\documentclass[10pt,prd,aps,twocolumn,a4papers,
floatfix,nofootinbib,showpacs,superscriptaddress]{revtex4-2}

\usepackage[utf8]{inputenc}
\usepackage{graphicx,psfrag}
\usepackage{graphics}
\usepackage[multidot]{grffile}

\usepackage{physics}
\usepackage{mathrsfs}          %
\usepackage{amsmath,amsfonts,amssymb}
\usepackage{multirow}
\usepackage{comment}
\usepackage{enumerate}
\usepackage{tabularx}

\usepackage{ragged2e}
\usepackage{subcaption} %
\DeclareCaptionJustification{justified}{\justifying}

\usepackage[colorlinks=true]{hyperref}
\hypersetup{
  linkcolor=blue,
  citecolor=blue,
  urlcolor=blue
}
\usepackage{cleveref}
\newcommand{\aref}[1]{\hyperref[#1]{App.~\ref*{#1}}}

\usepackage{bbm}
\usepackage{placeins}
\usepackage[normalem]{ulem}
\usepackage{tensor} %
\usepackage{xcolor}
\usepackage{orcidlink}

\usepackage{ragged2e} %
\DeclareCaptionJustification{justified}{\justifying}
\usepackage{verbatim}     %

\def\MADM{M_{\textrm{ADM}}}
\def\Mtot{M_{\textrm{tot}}}
\def\MBBS{M_{\textrm{BBS}}}
\def\MPl{M_{\textrm{Pl}}}
\def\TPl{T_{\textrm{Pl}}}
\def\LPl{L_{\textrm{Pl}}}
\def\hMPl{\hat{M}_{\textrm{Pl}}}
\def\hTPl{\hat{T}_{\textrm{Pl}}}
\def\hLPl{\hat{L}_{\textrm{Pl}}}
\def\PsiF{\Psi_{4}}
\def\PsiFT{\Psi_{4,20}}
\def\fcut{f_{\textrm{cut}}}
\def\Rext{r_{\textrm{ext}}}
\def\Nextrp{N_{\textrm{extrp}}}
\def\Cmon{\mathcal{C}_{\textrm{mon}}}

\def\Amax{A_{\textrm{max}}}
\def\bamps{\textsc{bamps}}
\def\sup{\textrm{sup}}
\newcommand{\eg}{e.\,g.\@}
\newcommand{\ie}{i.\,e.\@}
\newcommand{\vs}{vs.\@}

\definecolor{cyan}{rgb}{0,0.9,0.9}
\definecolor{orange}{rgb}{0.9,0.5,0}
\definecolor{purple}{rgb}{0.8,0.4,0.8}
\definecolor{gray}{rgb}{0.8242,0.8242,0.8242}
\definecolor{grey}{rgb}{0.5,0.5,0.5}
\definecolor{pink}{rgb}{1.0, 0.0, 0.5}

\newcommand{\mytext}[1]{}

\makeatletter
\newcommand\ID[2]{#1\def\@currentlabel{#1}\label{#2}}
\makeatother

\renewcommand{\mytext}{}

\newenvironment{datastatement}{%
  
  \begin{acknowledgments}
}{%
  \end{acknowledgments}
}

\begin{document}

\title{Boson star head-on collisions with constraint-violating\\and constraint-satisfying initial data}

\newcommand{\affjena}{Theoretisch-Physikalisches Institut, Friedrich-Schiller-Universit\"at Jena, 07743, Jena, Germany}
\newcommand{\afflisbon}{CENTRA, Departamento de F\'{\i}sica, Instituto Superior T\'ecnico -- IST, Universidade de Lisboa -- UL, Avenida Rovisco Pais 1, 1049-001 Lisboa, Portugal}
\newcommand{\affcambridge}{Department of Applied Mathematics and Theoretical Physics, Centre for Mathematical Sciences, University of Cambridge, Wilberforce Road, Cambridge CB3 0WA, United Kingdom}

\author{Florian \surname{Atteneder}\,\orcidlink{0000-0002-6617-5482}}
\affiliation{\affjena}

\author{Hannes R. \surname{Rüter}\,\orcidlink{0000-0002-3442-5360}}
\affiliation{\afflisbon}

\author{Daniela \surname{Cors}\,\orcidlink{0000-0002-0520-2600}}
\affiliation{\affjena}
\affiliation{\affcambridge}

\author{Roxana \surname{Rosca-Mead}\,\orcidlink{0000-0001-5666-1033}}
\affiliation{\affjena}

\author{David \surname{Hilditch}\,\orcidlink{0000-0001-9960-5293}}
\affiliation{\afflisbon}

\author{Bernd \surname{Brügmann}\,\orcidlink{0000-0003-4623-0525}}
\affiliation{\affjena}

\date{\today}

\begin{abstract}
  Simulations of binary collisions involving compact objects require initial data
  that satisfy the constraint equations of general relativity.
  For binary boson star simulations it is common practice to use a superposition
  of two isolated star solutions to construct an approximate solution to the constraint equations.
  Such superposed data is simple to set up compared to solving these equations explicitly,
  but also introduces extra constraint violations in the time evolution.
  In this work we investigate how physical observables depend on
  the quality of initial data in the case of head-on boson star collisions.
  In particular we compare results obtained from data prepared using
  four different methods: the standard method to superpose isolated stars,
  a heuristic improvement to this superposition technique
  and two versions of this data where excess constraint violations
  were removed through a conformal thin-sandwich solver.
  We find that differences in the time evolutions are dominated
  by differences in the way the two superposition methods differ,
  whereas additionally constraint solving the superposed data has smaller impact.
  The numerical experiments are conducted using the pseudospectral code \bamps{}.
  Our work demonstrates that \bamps{} is a code suited for generating high accuracy
  numerical waveforms for boson star collisions due to the exponential convergence in the polynomial resolution
  of the numerical approximation.
\end{abstract}

\maketitle

\section{Introduction}
\label{Section:Introduction}

With the first successful numerical relativity (NR) simulations of
binary black hole (BH) collisions carried out~\cite{pretorius2005evolution,
campanelli2006accurate,
baker2006gravitational,
brugmann2004numerical,
scheel2006solving},
an industry for engineering waveform templates was founded
which played a key role in the first gravitational wave (GW)
detections~\cite{abbott2016observation,abbott2016gw151226,abbott2016ligo}.
The late inspiral part of GW templates are informed by NR simulations
that solve an initial-boundary-value problem posed by a Cauchy formulation
of the Einstein field equations (EFEs) of general relativity (GR).
Essential to such simulations is the data described
on an initial hypersurface which is then propagated forward in time
to trace out a foliation of spacetime.
This initial data should satisfy the Hamiltonian and momentum
constraint equations of GR.
If the initial data also involve matter models in the form of neutron stars (NSs)
then one must also ensure that each star is initially in a state of quasi-equilibrium,
\ie{} this should account for effects like tidal deformations which are due to them
inspiraling on each other at a finite distance.
Given the difficulty of numerically solving the constraint equations to generate
such \textit{physically plausible} data a variety of
formalisms and numerical methods have been developed for collisions in the context
of BHs and NSs (see~\cite{tichy2016initial} for a review).

Almost a decade after the first GW detection the waveform template industry
continues pushing forward analytical, computational, and phenomenological boundaries in order to
keep up with the advancing detector technology and upcoming new experiments.
Among these theoretical advances is also considerable effort to study compact objects
described by exotic matter models that have been developed as dark matter candidates.
Among those candidates are boson stars (BSs), which were first theorized in~\cite{kaup1968klein},
and for which a variety of NR studies involving this particular model have been conducted.
These studies uncovered a dynamical formation process termed gravitational cooling~\cite{seidel1994formation},
as well as the anatomy of the GW signals from coalescences of such objects,
see~\cite{helfer2022malaise,evstafyeva2023unequal,croft2023gravitational,
palenzuela2007head}
for head-on collisions
and~\cite{palenzuela2008orbital,
bezares2017final,
bezares2022gravitational,
palenzuela2017gravitational,
sanchis2020synchronized}
for inspirals.
Also see~\cite{liebling2023dynamical,Mielke1997,Visinelli2021} for a review on BSs and variations thereof.

Most of today's understanding of BS collisions comes from studies that use a superposition of
isolated BSs as initial data which does not satisfy the constraint equations.
Such data should be seen as approximate solutions to the constraint equations.
Its use can be justified when it is explicitly verified
that the error due to initial constraint violations does not dominate the overall error budget
of the simulation.
Only recently, work has started on refining the initial data construction for BS encounters.
In~\cite{helfer2022malaise,evstafyeva2023unequal} a heuristically motivated
correction to the commonly used method of superposition was developed
by minimizing initial constraint violations, with some still remaining.
To the best of our knowledge the first BS collisions using constraint solved data were
reported in~\cite{dietrich2018full}, which discussed BS head-on encounters
only to calibrate the \textsc{BAM} code for evolutions of mixed NS-BS systems.
In~\cite{siemonsen2023generic} a generic constraint solver for BS initial data was developed,
which is an important contribution in order to catch up with the state-of-the-art
initial data construction techniques commonly used for BH and NS simulations.
Such constraint solved data was used to demonstrate that the collision
of two non-rotating BSs can yield a rotating remnant~\cite{siemonsen2023binary}.

As the sensitivity of the next generation of GW detectors increases, the accuracy demands
on GW templates do as well. In anticipation of such a trend new NR codes
are developed and existing NR codes are upgraded
to improve the computational efficiency and the mathematical modeling of physical processes, \eg{}
\textsc{BAM}~\cite{gonzalez2023second},
\textsc{GR-Athena++}~\cite{daszuta2021gr},
\textsc{Dendro-GR}~\cite{fernando2022massively},
\textsc{Einstein-Toolkit}~\cite{loffler2012einstein},
\textsc{ExaHyPE}~\cite{reinarz2020exahype},
\textsc{GRChombo}~\cite{clough2015grchombo},
\textsc{Lean}~\cite{sperhake2007binary},
\textsc{MHDuet}~\cite{bezares2022no},
\textsc{NMesh}~\cite{tichy2022new},
\textsc{NRPy+}~\cite{ruchlin2018senr},
\textsc{SACRA}~\cite{lam2023numerical},
\textsc{SpEC}~\cite{boyle2019sxs},
\textsc{SpECTRE}~\cite{kidder2017spectre},
\textsc{SPHINCS\_BSSN}~\cite{diener2022simulating},
\textsc{Spritz}~\cite{cipolletta2020spritz}
(see also \cite{afshordi2023waveform} for an extended list).
Among these research efforts is also the \bamps{} code~\cite{hilditch2016pseudospectral,
brugmann2013pseudospectral,
bugner2016solving,
bhattacharyya2021implementation,
ruter2018hyperbolic,
renkhoff2023adaptive,
hilditch2017evolutions,
fernandez2022evolution,
cors2023formulation}
which employs a nodal pseudospectral (PS) discretization for the spatial representation
of the solution.
The promised efficiency and accuracy gain of a PS method can be demonstrated
best for problems which admit smooth solutions.
Since BSs lack a hard surface (or boundary) and no shock fronts are formed during
mergers of such objects, which are common obstacles for BH and NS simulations that spoil smoothness,
they represent an ideal testbed to develop and assess PS methods in the context of GR.
Theoretically, the solution has exponential convergence when increasing
the polynomial resolution of the PS approximations.
This translates into a significantly reduced error budget that we attribute to the time evolution,
assuming high enough resolution.
Thus, for PS codes constraint violations
in superposed initial data can in principle dominate the total error budget of our results, making
the usage of superposed initial data ill-advised.

In this work we utilize the \bamps{} code to perform binary BS head-on collisions
with axisymmetry and reflection symmetry and investigate how physical observables
extracted from these simulations depend on the quality of initial data.
In particular, we compare results obtained from data that was prepared using
four different methods: a simple superposition of isolated stars
as defined in~\cite{helfer2022malaise},
the heuristic improvement to the simple superposition technique
also reported in~\cite{helfer2022malaise}, and two versions of constraint-solved data
obtained from a conformal thin-sandwich (CTS) solver and superposed free data.
We then assess the differences between evolutions done with these data
and the individual accuracy of each evolution by using a mixture
of constraint monitors, global (and conserved) physical quantities,
as well as gravitational waves, and (self-)convergence tests involving those quantities.

The rest of this work is structured as follows.
First, we review the theory under study as well as the formulation of the equations of motion
we use for the numerical simulations in \autoref{Section:Theory}.
In \autoref{Section:InitialData} a summary about earlier work on superposed
initial data is given and we discuss how we construct constraint-solved data for
the comparisons.
Details on the computational setting are provided in \autoref{Section:setup} and
our numerical results are presented in \autoref{Section:Results}.
A summary of our findings is provided in \autoref{Section:Conclusion}.
In \aref{Appendix:gw-analysis} and \aref{Appendix:gw-wiener-product} we discuss details of our GW analysis and
comment on a problem specific to head-on collisions and the reconstruction of the GW strain $h$
from the Newman-Penrose pseudoscalar $\Psi_4$.
We use Latin indices starting from $a$ to denote
spacetime components and Latin indices starting from $i$ to denote spatial components
of tensors.
We work in Planck units where $G = c = \hbar = 1$ so that all variables are
automatically dimensionless and we also set the
scalar-field mass (defined below) to $\mu = 1$.
In \aref{Appendix:units} we show that this choice of $\mu$ does not limit the
generality of our results, but instead corresponds to
a particular rescaling of the variables.

\section{Theory}
\label{Section:Theory}

\subsection{Action and equations of motion}

This work is concerned with scalar BSs in GR. They are compact objects
defined as solutions to the equations
of motion in a theory in which the Einstein-Hilbert action for the gravitational field $g_{ab}$
is minimally coupled to a complex scalar field $\phi$ in the following
way~\cite{liebling2023dynamical}
\begin{align}
  S = \int &\dd^4 x~\sqrt{-g} \nonumber\\
      &\times \left(
      \frac{\tensor[^{(4)}]{R}{}}{16\pi}
      - \frac{1}{2} \left( g^{ab} \nabla_a \phi^\ast \nabla_b \phi + V(|\phi|^2) \right)
  \right) \, ,
  \label{eq:action}
\end{align}
where $g$ is the metric determinant, $\tensor[^{(4)}]{R}{}$
the Ricci scalar associated with $g_{ab}$ and $V(|\phi|^2)$ is the scalar-field potential,
and an asterisk refers to complex conjugation.
The above action gives rise to the following equations of motion,
known as the Einstein-Klein-Gordon (EKG) equations,
\begin{align}
  G_{ab} &= 8 \pi T_{ab} \, ,
  \label{eq:efe}
  \\
  \Box \phi &= \phi \frac{\dd V}{\dd |\phi|^2} \, ,
  \label{eq:kleingordon}
\end{align}
where $G_{ab}$ is the Einstein tensor, $\Box \equiv g^{ab} \nabla_{a} \nabla_{b}$
and the stress-energy tensor is given by
\begin{align}
  T_{ab} &= \nabla_{(a} \phi^\ast \nabla_{b)} \phi
    - \frac{1}{2} g_{ab} \left(
      g^{cd} \nabla_{c} \phi^\ast \nabla_{d} \phi + V(|\phi|^2)
    \right) \, .
  \label{eq:stress-energy-tensor}
\end{align}

BSs can come in different flavors determined by the form of the potential $V(|\phi|^2)$.
In this work we restrict ourselves to mini BSs for which the potential is that
of a massive free scalar field,
\begin{align}
  V(|\phi|^2) = \mu^2 |\phi|^2 \, ,
  \label{eq:potential-free-scalar}
\end{align}
where $\mu$ is the scalar field's mass. For the rest of this work we set $\mu = 1$.

\subsection{3+1 decomposition}

The basis for NR simulations is laid by a covariant 3+1 decomposition of the
spacetime metric $\tensor{g}{_a_b}$, often written in the form~\cite{baumgarte2010numerical}
\begin{align}
  g_{ab} \dd x^a \dd x^b = - \alpha^2 \dd t^2 + \gamma_{ij} (\dd x^i + \beta^i \dd t)
  (\dd x^j + \beta^j \dd t) \, .
  \label{eq:3+1-metric}
\end{align}
The variables $\alpha$, $\beta^i$, $\gamma_{ij}$ and $K_{ij}$ are called the lapse, shift, spatial metric
and extrinsic curvature, respectively, and are referred to as the 3+1 variables.
Within this framework the EFEs are rewritten accordingly and this unveils the so-called
Hamiltonian and momentum constraint equations (from here on only referred to as constraint equations)
as part of this system of nonlinear PDEs~\cite{baumgarte2010numerical},
\begin{align}
  \mathcal{H} &:= R + K^2 - K_{ij} K^{ij} - 16 \pi \rho = 0 \, ,
  \label{eq:hamilton-constraint}
  \\
  \mathcal{M}^{i} &:= D_j ( K^{ij} - \gamma^{ij} K ) - 8 \pi S^i = 0 \, ,
  \label{eq:momentum-constraint}
\end{align}
where $R$ and $K_{ij}$ are the Ricci scalar and extrinsic curvature associated with $\gamma_{ij}$
and a spatial hypersurface $\Sigma$ that is embedded in the surrounding spacetime
$(\mathcal{M},\tensor{g}{_a_b})$.
The quantities $\rho = n^a n^b T_{ab}$ and $S^i = - \gamma^{ia} n^b T_{ab}$ are projections
of the stress-energy tensor $T_{ab}$
onto $(\Sigma,\tensor{\gamma}{_i_j})$,
where $n^a$ is the unit normal vector to $\Sigma$
and $\gamma^{ia} = g^{ia} + n^i n^a$.
These equations \textit{constrain} the fields
$(\gamma_{ij}, K_{ij})$ of a time slice $\Sigma$ such that the embedding
of $(\Sigma,\gamma_{ij},K_{ij})$ in $(\mathcal{M},\tensor{g}{_a_b})$ is compatible with the covariant
decomposition of the EFEs.

The remainder of the decomposition of the EFEs is complemented by how
$\gamma_{ij}$ and $K_{ij}$ develop away from the initial hypersurface $\Sigma_0$.
Augmenting these equations with conditions on how the variables $\alpha$ and $\beta^i$ evolve
completes the system of nonlinear PDEs we aim to solve.
In this work we utilize the generalized harmonic gauge (GHG) formulation of the EFEs~\cite{hilditch2016pseudospectral,lindblom2006new}
\begin{align}
\label{eq:ghg}
  \partial_t \tensor{g}{_a_b} &=
  \beta^i \partial_i \tensor{g}{_a_b}
    - \alpha \tensor{\Pi}{_a_b}
    + \gamma_1 \beta^i \tensor{C}{_i_a_b} \,,\\
\label{eq:ghg2}
  \begin{split}
    \partial_t \tensor{\Pi}{_a_b} &=
      \beta^i \partial_i \tensor{\Pi}{_a_b}
      - \alpha \gamma^{ij} \partial_i \tensor{\Phi}{_j_a_b}
      + \gamma_1 \gamma_2 \beta^i \tensor{C}{_i_a_b} \\
    &\quad + 2 \alpha g^{cd}
      \left( \gamma^{ij} \tensor{\Phi}{_i_c_a} \tensor{\Phi}{_j_d_b}
      - \tensor{\Pi}{_c_a} \tensor{\Pi}{_d_b}
      - g^{ef} \tensor{\Gamma}{_a_c_e} \tensor{\Gamma}{_b_d_f} \right) \\
    &\quad - 2 \alpha \left( \tensor{\nabla}{_(_a} \tensor{H}{_b_)}
      + \gamma_4 \tensor{\Gamma}{^c_a_b} \tensor{C}{_c}
      - \frac{1}{2} \gamma_5 \tensor{g}{_a_b} \Gamma^c \tensor{C}{_c} \right) \\
    &\quad -\frac{1}{2} \alpha n^c n^d \tensor{\Pi}{_c_d} \tensor{\Pi}{_a_b}
      - \alpha n^c \gamma^{ij} \tensor{\Pi}{_c_i} \tensor{\Phi}{_j_a_b} \\
    &\quad + \alpha \gamma_0 \left( 2 \tensor{\delta}{^c_{(a}} \tensor{n}{_{b)}}
      - \tensor{g}{_a_b} n^c \right) \tensor{C}{_c}\\
    &\quad - 16 \pi \alpha
      \left (T_{ab} - \frac{1}{2} g_{ab} \tensor{T }{^c_c} \right) \,,
  \end{split}\\
\label{eq:ghg3}
  \begin{split}
    \partial_t \tensor{\Phi}{_i_a_b} &=
      \beta^j \partial_j \tensor{\Phi}{_i_a_b}
      - \alpha \partial_i \tensor{\Pi}{_a_b}
      + \gamma_2 \alpha \tensor{C}{_i_a_b} \\
    &\quad + \frac{1}{2} \alpha n^c n^d
      \tensor{\Phi}{_i_c_d} \tensor{\Pi}{_a_b}
      + \alpha\gamma^{jk} n^c \tensor{\Phi}{_i_j_c}
      \tensor{\Phi}{_k_a_b} \,,
  \end{split}
\end{align}
where the evolved variables are the metric $g_{ab}$ and
the time reduction variable $\Pi_{ab} = -n^c \partial_c g_{ab}$,
$\Gamma^a = g^{bc} \tensor{\Gamma}{^a_b_c}$ and $\Gamma_{abc} = g_{ad} \tensor{\Gamma}{^d_b_c}$
are the Christoffel symbols associated with $g_{ab}$.
We would like to point out the extra minus sign in the definition of $\Pi_{ab}$
which was previously missing in~\cite{renkhoff2023adaptive} due to a typo.
The spatial reduction variable $\Phi_{iab}$ is associated with
the reduction constraint $C_{iab}=\partial_ig_{ab}-\Phi_{iab} = 0$ and
\begin{align}
   C_a = H_a + \Gamma_a = 0
   \label{eq:harmonic-gauge-constraint}
\end{align}
is the harmonic constraint, where $H_a$ is a gauge source function.
Here we choose the gauge source function introduced in~\cite{deppe2019critical}
with $R(t)=W(x^i)=1$.
The constraint damping parameters are fixed to be
$\alpha\gamma_0=1/10$, $\gamma_1=-1$, $\alpha\gamma_2=1$ and $\gamma_4=\gamma_5=1/2$.
The above evolution equations are implemented in our numerical relativity code \bamps{} and more
details on the computational setup are discussed in \autoref{Section:setup}.

To reduce the Klein-Gordon equation~\eqref{eq:kleingordon} to first order
we introduce the reduction variables
$\Pi = n^a\partial_a \phi$, $\Phi_i=\partial_i \phi$
and the spatial reduction constraint $B_i :=\partial_i \phi-\Phi_i$.
The reduced system of equations is then of the form
\begin{align}
    \partial_t \phi &= \alpha \Pi + \beta^i \Phi_i \, ,
    \label{eq:kleingordon-reduction1}
    \\
    \partial_t \Pi &=  \beta^i \partial_i \Pi
    \label{eq:kleingordon-reduction2}
    \\
    &+ \gamma^{ij}\left(\Phi_j\partial_i\alpha+\alpha\partial_i \Phi_j -\alpha {}^{(3)}\Gamma^k{}_{ij}\Phi_k\right) \nonumber\\
    &+\alpha\Pi K +\sigma \beta^i B_i \, , \nonumber \\
    \partial_t \Phi_i &=\Pi \partial_i\alpha+ \alpha\partial_i \Pi+\Phi_j\partial_i\beta^j
    \label{eq:kleingordon-reduction3}
    \\
    &+\beta^j\partial_j \Phi_i+\sigma\alpha B_i \nonumber \, .
\end{align}
The evolved variables are $\phi$, $\Phi_i$ and $\Pi$.
$^{(3)}\tensor{\Gamma}{^k_i_j}$ refer to the Christoffel symbols associated with
$\gamma_{ij}$ and $\sigma$ is a damping parameter which we set
in all our simulations such that $\alpha \sigma = 1$,
equivalent to our treatment of $\gamma_2$.

\section{Initial data}
\label{Section:InitialData}

\subsection{Binary boson star initial data from superposition of isolated stars}
\label{subsection:superposed-initial-data}

The construction of initial data for simulations of compact binary objects requires
finding a solution $(\gamma_{ij}, K_{ij})$
to the geometric constraints~\eqref{eq:hamilton-constraint} and~\eqref{eq:momentum-constraint}
on an initial time slice $\Sigma_0$,
while simultaneously also preparing the matter variables $\rho$ and $S^i$ in
a state of quasi-equilibrium determined through auxiliary conditions.
In the following discussion we neglect the latter aspect and focus only on the geometric
constraints, but we revisit it briefly below.
What defines \textit{physically plausible} initial data is in general
not a question with a simple answer.
The basis for binary data construction is the existence of isolated star solutions,
which are often assumed to be at least axisymmetric and, thus,
simple to obtain numerically.
Unfortunately, the assumption of a star being in isolation is, in principle, in contradiction
with a star partaking in a binary collision, unless they are displaced by an infinite distance.
On a physical basis this is clear, because the gravitational pull of one star will be felt
by its companion, causing tidal deformations of it and, hence, influencing
its gravitational potential, which in turn modifies the initial star through its own tidal forces.
This is also reflected by the nonlinearity of the constraint equations which in general
prevents the construction of new solutions as simple combinations of isolated single star solutions
-- commonly referred to as \textit{superposition}.

As can be straightforwardly shown, the constraint equations, if satisfied
at one instant of time and evolved exactly, will continue to hold at all times.
Failing to satisfy these constraints does not necessarily cause crashes in
numerical simulations, but results that do so are not solutions to the EFEs.
Realizing that NR can only ever produce approximate numerical solutions to the continuum EFEs,
any numerical result will inevitably violate the constraints to some degree.
The goal of NR is to construct successive numerical approximations
in such a way that constraint violations, as well as the evolved fields and other
analysis quantities, show a controlled convergence trend towards
the continuum EFEs in the limit of infinite resolution.
We emphasize that violations occurring in time evolutions do not solely arise
from the failure of the initial data to satisfy the constraints,
but can \eg{} also grow dynamically through numerical errors due to insufficient resolution,
the appearance of (coordinate) singularities or other pathologies.
In fact, constraint violations associated with the initial data quality are
conceptually even simpler to control,
because the mathematical problem of solving the constraint equations for the initial data
is completely decoupled from solving the free evolution equations.
Because of that, dedicated codes have been developed to tackle the initial data problem in GR
(see~\cite{tichy2016initial} for a review of methods for BH and NS binaries,
see~\cite{siemonsen2023binary} for a constraint solver for BS binaries).

Solving the constraint equations is a difficult task.
Using instead a superposition of isolated single star solutions
presents itself as an attractive shortcut to constructing
initial data that solve the constraints approximately\footnote{
  We refer to~\cite{helfer2022malaise} for a brief list of special cases in which
  the superposition of solutions can solve the constraint equations exactly.
}.
In studies that use such data it is often argued that the approximation is accurate
enough when the inherited constraint violations are at most of the same order as
the error budget of the time evolutionary part of a numerical code~\cite{palenzuela2008orbital}.
Such an argument can be supported with convergence studies to demonstrate that the numerical
results approximate solutions up to a controlled error,
provided that one commits to routinely verify that the above premise is satisfied.
In the case of BS simulations superposed data has been used many times
before as starting points for head-on and inspiraling binary
collisions~\cite{palenzuela2008orbital,
palenzuela2007head,
bezares2017final,
bezares2022gravitational,
palenzuela2017gravitational,
sanchis2020synchronized,
helfer2022malaise,
evstafyeva2023unequal,
croft2023gravitational}.
The same technique has also been employed for collisions involving other exotic matter models like
Proca stars~\cite{sanchis2022impact,
bustillo2021gw190521,
sanchis2019head},
$\ell$-boson stars~\cite{jaramillo2022head},
dark boson stars~\cite{bezares2018gravitational},
axion stars~\cite{sanchis2022electromagnetic},
neutron stars with bosonic cores~\cite{bezares2019gravitational}
as well as mixed mergers of an
axion star and a black hole~\cite{clough2018axion},
a neutron star and an axion star~\cite{dietrich2019neutron, clough2018axion},
and a boson star and a black hole~\cite{cardoso2022piercing,zhong2023piercing}.

Focusing on pure scalar BS simulations, one recipe employed in the literature to
construct superposed data can be summarized as follows.
Let $\gamma_{ij}^A$, $K_{ij}^A$, $\alpha_A$ and $\beta_A^i$ denote the 3+1 variables of spacetime
and let $\phi_A$ and $\Pi_A$ denote the scalar-field variables of star $A$,
likewise for star $B$.
A \textit{simple superposition} (SSP) initial data construction involving
stars $A$ and $B$ is then given by~\cite{helfer2022malaise}
\begin{align}
  \gamma_{ij} &= \gamma_{ij}^{A} + \gamma_{ij}^B - \delta_{ij} \, ,
  \label{eq:naive-superposition-gamma}
  \\
  K_{ij} &= \gamma_{m(i} \left(
    K_{j)n}^{A} \gamma_{A}^{nm} + K_{j)n}^{B} \gamma_{B}^{nm}
  \right) \, ,
  \label{eq:superposition-K}
  \\
  \phi &= \phi_A + \phi_B \, ,
  \label{eq:naive-superposition-phi}
  \\
  \Pi &= \Pi_A + \Pi_B \, ,
  \label{eq:naive-superposition-pi}
\end{align}
where the data of the isolated stars is boosted such that
the superposition mimics a binary configuration where each star carries initial momenta.
This boosting itself is readily done through Galilean or Lorentz boost transformations~\cite{palenzuela2008orbital,helfer2022malaise}.
A slight variation of this recipe has been used in other studies
that fueled many important contributions to the
field~\cite{palenzuela2008orbital,
palenzuela2007head,
bezares2017final,
bezares2022gravitational,
palenzuela2017gravitational}.

The above construction disregards the constraint equations.
Consequently, it comes as no surprise that in~\cite{helfer2022malaise}
it has been demonstrated that the scalar-field amplitude can admit artificial modulations.
These effects can be strong enough to trigger premature BH collapse in encounters
of BSs with a solitonic potential.
Similar results were reported earlier in~\cite{helfer2019gravitational} for the
case of compact real-scalar solitons (oscillatons).
To tackle these symptoms, the same authors came up
with a heuristically motivated improvement to the SSP construction, guided
by minimizing initial constraint violations,
which we refer to as the \textit{constant volume element} (CVE) construction.
Since the ansatz~\eqref{eq:naive-superposition-gamma} alters the volume form
inside a star depending on the displacement of the companion star, it was adjusted to
\begin{align}
  \gamma_{ij} &= \gamma_{ij}^{A} + \gamma_{ij}^B - \gamma_{ij}^{A}(x_{B}) \, .
  \label{eq:helfer-superposition-gamma}
\end{align}
Here $\gamma_{ij}^A(x_{B})$ refers to the constant value of the components of the induced metric
of star A, $\gamma_{ij}^A$, evaluated at the center of star B, $x_B$.
Note that in the case of equal mass BS binary systems
we have $\gamma_{ij}^{A}(x_{B}) = \gamma_{ij}^{B}(x_{A})$.
Consequently, the above ensures that $\gamma_{ij}(x_A) = \gamma_{ij}^A(x_A)$ and likewise for B,
hence, it approximately restores the volume element form around each star's center to the value
of an isolated star.
This simple correction was enough to cure premature BH collapse,
but at the same time, the correction led to qualitative changes in the emitted GW
signals~\cite{helfer2022malaise}.
Recently, \cite{evstafyeva2023unequal}~generalized the CVE construction to also
work with unequal mass binaries.

\begin{figure}
  \pgfimage[width=0.95\columnwidth]{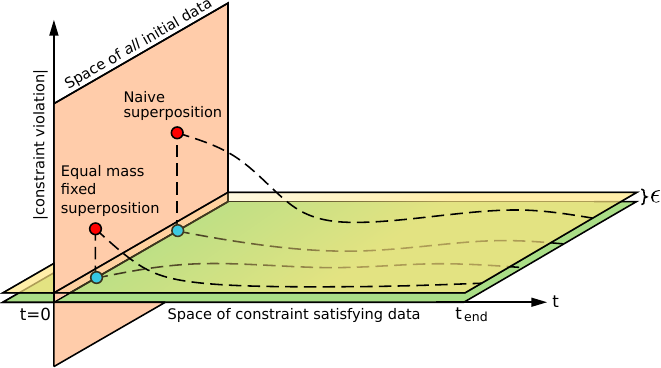}
  \caption{
  Schematic representation of the space of initial data (orange) and of
  constraint-satisfying data (green).
  The red dots correspond to superposed initial data constructed with the SSP or CVE technique.
  The blue dots correspond to a projection of SSP and CVE data
  (in this work we use a CTS constraint solver for this, see text) onto the space of
  constraint-satisfying data.
  Dashed lines indicate the trajectories of the data when evolved through an
  evolution system with a damping scheme.
  Initially constraint-violating data might end up a distance $\epsilon$ displaced (yellow)
  and never reach the space of constraint-satisfying data.
  }
  \label{fig:sketch-constraints}
\end{figure}

We now have two methods at our disposal to construct superposed initial data
and one might ask which of those is preferred, given that their time evolutions
can yield different physical results.
One might be inclined to prefer the CVE construction, because it reduces
constraint violations and cures premature BH collapse.
But whether these differences are only caused by the improvement of the constraint violations
or are perhaps primarily due to changes in physical characteristics of the initial data,
like for instance the local energy density, is yet unclear.
To further illustrate this point consider \autoref{fig:sketch-constraints} which
schematically shows how constraint violations (measured in some norm) associated
with initial data might propagate in time, when using an evolution system
with a built-in constraint damping scheme like GHG.
The CVE construction provides initial data
with less constraint violations than the initial data constructed with
the SSP technique, which is indicated by the two
red dots lying on the space of all initial data (orange) at $t=0$
while also being displaced vertically from the space of constraint-satisfying data (green).
The two blue dots, on the other hand, represent two different initial data sets
(which are constructed in this work)
where all excess constraint violations (ignoring numerical error)
were removed from the superposed initial data.
Performing time evolutions will then trace out the dashed trajectories,
which indicate that constraint-satisfying initial data
remain constraint-satisfying throughout the evolution.
Contrary to this, evolutions starting from
superposed initial data will only gradually in time approach the space of constraint-satisfying data.
Note that the seemingly attractive character of the space of constraint-satisfying data is usually due to two factors:
1) the use of modified evolution equations that include constraint damping terms
and 2) the possibility that constraint violations can propagate and eventually leave
the computational domain through a boundary.
Furthermore, there is no guarantee that initially-constraint-violating data is going to
converge with increasing resolution towards constraint-satisfying data at late times.
Instead one should expect some excess violations to remain also for late times $t$,
which is illustrated by the trajectories ending up on the yellow space that
is displaced by $\epsilon$ from the space of constraint-satisfying data.
Besides this vertical displacement at late times, it is also unclear whether
two trajectories that emanated from two initial data sets that are based on the same superposition,
but where one is constraint-violating and one constraint-satisfying,
will end up close to each other.
Below we investigate the qualitative behavior of the sketched trajectories for the case of selected
binary BS configurations and study their behavior when varying numerical resolution.

The above discussion as well as the rest of this work does not account
for the problem of the progenitors not being initially in a state of quasi-equilibrium,
which is done to ease this comparison.
We want to highlight that the constraint equations need to be satisfied
regardless of the matter model considered to study solutions to the EFEs,
and matter fields enter these equations only as source terms.
On the other hand, the equations of motion of most matter models, including BSs,
do not provide any constraints, meaning that, in principle, arbitrary matter configurations
could be used as initial data, provided the metric fields are adjusted to satisfy
the geometric constraints.
Instead, one requires additional assumptions to define a quasi-equilibrium, \eg{}
the existence of a helical Killing vector field for binary inspiral configurations,
as well as an understanding of the matter model to derive (elliptic) equations that
equilibrate their fields accordingly.
Special attention has been given to the latter aspect of initial data construction in
the BH and NS literature, see~\cite{cook2000initial,
pfeiffer2005initial,
gourgoulhon2007construction,
baumgarte2010numerical}.
First work in this direction for binary BSs has started only recently in~\cite{siemonsen2023generic}
and we leave it to future work to further study the importance of this facet.

In the next section we examine
one possibility to remove excess constraint violations
from SSP and CVE initial data by numerically solving the constraint equations using
a CTS solver, where the free data is constructed from the SSP and CVE data.
Subsequently, we perform time evolutions using this data in order to
answer the question of whether differences in physical observables between SSP and
CVE initial data are caused by constraint violations. Along the way we also
conduct convergence studies to assess the quality of our numerical results.

\subsection{Constraint satisfying binary boson star initial data}

Constraint satisfying binary BS initial data has been constructed before
in~\cite{dietrich2018full}
as well as in~\cite{siemonsen2023binary,siemonsen2023generic}
and was an important ingredient in demonstrating that the collision of two non-rotating BSs
can form a rotating remnant. Below we follow a similar procedure by using
the CTS formulation of the Hamiltonian and momentum
constraints~\eqref{eq:hamilton-constraint} and~\eqref{eq:momentum-constraint},
which read~\cite{york1999conformal}
\begin{align}
  \bar{D}^2 \psi &- \frac{1}{8} \psi \bar{R} - \frac{1}{12} \psi^5 K^2
  + \frac{1}{8} \psi^{-7} \bar{A}_{ij} \bar{A}^{ij}
  \nonumber
  \\
  &= - 2 \pi \psi^5 \rho \, ,
  \label{eq:cts-psi}
  \\
  ( \bar{\Delta}_L \beta)^i &- (\bar{L}\beta)^{ij} \bar{D}_j \log(\bar{\alpha})
  \nonumber
  \\
  &= \bar{\alpha} \bar{D}_j (\bar{\alpha}^{-1} \bar{u}^{ij})
  + \frac{4}{3} \bar{\alpha} \psi^6 \bar{D}^i K
  \nonumber
  \\
  &+ 16 \pi \bar{\alpha} \psi^{10} S^i \, ,
  \label{eq:cts-beta}
\end{align}
where
\begin{align}
  \bar{A}^{ij} = \frac{1}{2 \bar{\alpha}} ( (\bar{L} \beta)^{ij} - \bar{u}^{ij}) \, .
\end{align}
The above constitute a system of four elliptic PDEs for the conformal factor $\psi$
and the components of the shift vector $\beta^i$, where
$\bar{D}$ is the covariant derivative associated with the conformal
metric $\bar{\gamma}_{ij}$, $\bar{R}$ is the Ricci scalar associated with $\bar{\gamma}_{ij}$
and~\cite{baumgarte2010numerical}
\begin{align}
  (\bar{\Delta}_L \beta)^i &= \bar{D}^2 \beta^i + \frac{1}{3} \bar{D}^i( \bar{D}_j \beta^i )
  + \tensor{\bar{R}}{^i_j} \beta^j \, ,
  \\
  (\bar{L} \beta)^{ij} &= \bar{D}^i \beta^j + \bar{D}^j \beta^i - \frac{2}{3} \bar{\gamma}^{ij} \bar{D}_k \beta^k \, ,
\end{align}
are the conformal versions of the vector Laplacian and vector gradient applied to $\beta^i$,
respectively.
Once a solution for $\psi$ and $\beta^i$ is known,
the remaining 3+1 variables can be recovered from
\begin{align}
  \gamma_{ij} &= \psi^4 \bar{\gamma}_{ij} \, ,
  \\
  K_{ij} &= \psi^{-2} \bar{A}_{ij} + \frac{1}{3} \gamma_{ij} K \, ,
  \\
  \alpha &= \psi^6 \bar{\alpha} \, .
  \label{eq:cts-alpha}
\end{align}

Equations~\eqref{eq:cts-psi} and~\eqref{eq:cts-beta} need to be complemented with freely specifiable data
for $\bar{\gamma}_{ij}$, its time derivative $\bar{u}_{ij} := \partial_t \bar{\gamma}_{ij}$,
the trace of the extrinsic curvature $K$
and the conformal lapse $\bar{\alpha}$.
Because we want to test the influence of eliminating constraint violations
from superposed initial data, denoted by $(\gamma_{ij}^{(\sup)}, K^{(\sup)})$,
we use that data to setup the free data for the CTS formulation, following~\cite{east2012conformal},
\begin{align}
  \bar{\gamma}_{ij} &= \gamma_{ij}^{(\sup)} \, ,
  \label{eq:sup-gamma}
  \\
  \bar{u}_{ij} &= u_{ij}^{(\sup)} \, ,
  \\
  \bar{u}^{ij} &= \bar{\gamma}^{ik} \bar{\gamma}^{jl}
  \left(
    \bar{u}_{kl}^{(\sup)} - \frac{1}{3} \bar{\gamma}^{mn} \bar{u}_{mn} \bar{\gamma}_{kl}
  \right) \, ,
  \label{eq:ubar}
  \\
  K &= K^{(\sup)} \, ,
  \\
  \bar{\alpha} &= \alpha^{(\sup)} = \det(\gamma_{ij}^{(\sup)})^{-1/6} \, ,
  \label{eq:lapse-sup}
  \\
  {\beta^{(\sup)i}} &= 0 \, ,
  \label{eq:shift-sup}
\end{align}
and
\begin{align}
  u_{ij}^{(\sup)} =
  \partial_t \gamma_{ij}^{(\sup)} =
  - 2 \alpha^{(\sup)} K^{(\sup)} + \mathcal{L}_{\boldsymbol{\beta}^{(\sup)}} \gamma_{ij}^{(\sup)} \, ,
  \label{eq:sup-uij}
\end{align}
where $\mathcal{L}_{\boldsymbol{\beta}^{(\sup)}}$ denotes the Lie transport along
the shift vector $\beta^{(\sup)i}$ of the superposed data.
The choices~\eqref{eq:lapse-sup} and~\eqref{eq:shift-sup} were adopted
from~\cite{helfer2022malaise}.
Note that the difference in sign in~\eqref{eq:ubar} when compared to~\cite{east2012conformal}
is due to a difference in notation (see \aref{Appendix:notation-conformal-metric}).

The metric variables
$\gamma_{ij}^{(\sup)}$, $K^{(\sup)}$, $\alpha^{(\sup)}$ and $\beta^{(\sup)i}$ are given
by~\eqref{eq:naive-superposition-gamma}, \eqref{eq:superposition-K}, \eqref{eq:lapse-sup} and~\eqref{eq:shift-sup}
or~\eqref{eq:helfer-superposition-gamma}, \eqref{eq:superposition-K}, \eqref{eq:lapse-sup} and~\eqref{eq:shift-sup}, respectively.
The matter source terms $\rho$ and $S^i$
are computed from $\gamma_{ij}^{(\sup)}$, $K^{(\sup)}$, $\alpha^{(\sup)}$, $\beta^{(\sup)}$
and $T_{ab}$ given by~\eqref{eq:stress-energy-tensor}.
The construction of the latter is done by using the metric variables as well as the matter variables
$\phi$ and $\Pi$ as given by~\eqref{eq:naive-superposition-pi}
and~\eqref{eq:helfer-superposition-gamma}.
Note that, although the way the scalar field is superposed
by~\eqref{eq:naive-superposition-phi} and~\eqref{eq:naive-superposition-pi}
is the same between the SSP and CVE method,
the difference in the superposition of the induced metric $\gamma_{ij}$, given by~\eqref{eq:naive-superposition-gamma} and~\eqref{eq:helfer-superposition-gamma},
then also
translates into differences in the stress-energy tensor between these two constructions.

The numerical solution of~\eqref{eq:cts-psi} and~\eqref{eq:cts-beta}
is obtained using the hyperbolic relaxation method~\cite{ruter2018hyperbolic},
which is available inside the \bamps{} code.
For the CTS solver we use Robin boundary conditions to impose the asymptotic
behavior~\cite{ruter2018hyperbolic}
\begin{align}
  \alpha = \psi  &= 1 + \mathcal{O}(r^{-1}) \, ,
  &
  \beta^i &= \mathcal{O}(r^{-1}) \, .
\end{align}
To set up the free data for the solver we proceed as follows:
first, we compute spherically symmetric stationary solutions of isolated BSs;
two such stars are then each boosted by a parameter $v$ using a Lorentz transformation.
For both of these steps we follow closely the algorithms given in~\cite{helfer2022malaise}
(also see \aref{Appendix:isolated-bs}).
We then use these stars to construct SSP or CVE initial data
following the algorithms given in~\autoref{subsection:superposed-initial-data}.
These SSP and CVE data are then used together with~\eqref{eq:sup-gamma}-\eqref{eq:sup-uij}
to set up the free data and sources for solving the CTS
equations~\eqref{eq:cts-psi} and~\eqref{eq:cts-beta}.
As an initial guess for the CTS solver we use
\begin{align}
  \psi &= 1 \, ,
  &
  \beta^i &= 0 \, .
  \label{eq:cts-initial-guess}
\end{align}
The hyperbolic relaxation of the CTS solver terminates once
the sum of the $L^1$ norm of the right-hand side (RHS) of equations~\eqref{eq:cts-psi} and~\eqref{eq:cts-beta}
is smaller than $1/10$th the sum of the $L^1$ norm of the residuals of those equations,
or when the $L^1$ norm of the residual of the equations falls below $10^{-8} \times \#N_{\text{DOF}} \times 4$,
where $N_{\text{DOF}}$ is the total number of grid points on the target resolution.
The resulting constraint-satisfying data is referred to as
CTS+SSP and CTS+CVE data for the rest of this work.

When we consider evolutions of constraint-satisfying data below, for which we fix
the grid structure and the polynomial resolution $n$ in each cell,
but we vary $n$ between different runs, the initial data is computed using
the hyperbolic relaxation method on the very same grid structure
and with the same polynomial resolution.
In particular, no extra interpolation step is needed to convert the solution of the \bamps{}
internal CTS solver into initial data for the evolution.

For evolutions of superposed data we utilize~\eqref{eq:lapse-sup} and~\eqref{eq:shift-sup}
as initial data for $\alpha$ and $\beta^i$.
For the case of constraint-solved data we use instead the solutions obtained from the CTS equations~\eqref{eq:cts-psi},
\eqref{eq:cts-beta} and the relation~\eqref{eq:cts-alpha}
to set up $\alpha$ and $\beta^i$.

\section{Computational setup}
\label{Section:setup}

\begin{table}[t]
 \centering
 \caption{
   Summary of the grid configuration used for all the results presented in this work.
   The highest resolved configuration ($n=21$ and 48 outer shells) uses $423360$ degrees of freedom
   (already accounting for axisymmetry and reflection symmetry).
 }
  \begin{tabularx}{\columnwidth}{XXX}
    \hline
      Parameter & value & description \\
    \hline
      \texttt{grid.cube.max}                     & 80        & 1/2 side length of inner cube \\
      \texttt{grid.sub.xyz}                      & 16        & number of subdivisions in inner cube \\
      \texttt{grid.cubedsphere} \texttt{.max.x}  & 160       & outer radius of cube-to-sphere patch \\
      \texttt{grid.cubedsphere} \texttt{.sub.x}  & 8         & number of radial subdivisions in cube-to-sphere patch \\
      \texttt{grid.sphere} \texttt{.max.x}       & [400,800] & outer radius of sphere path \\
      \texttt{grid.sphere} \texttt{.sub.x}       & [24,48]   & number of radial subdivisions in sphere patch \\
      \texttt{grid.dtfactor}                     & 0.25      & CFL factor \\
      \texttt{grid.cartoon}                      & xz        & double cartoon method \\
      \texttt{grid.reflect}                      & z         & reflection symmetry across $z=0$ plane \\
      \texttt{grid.n.xyz}                        & $n :=$ [7,9,11,13,\newline
                                                   15,17,19,21]  & number of points per grid and per dimension \\
    \hline
  \end{tabularx}
 \label{tab:grid-parameters}
\end{table}

\bamps{}~\cite{brugmann2013pseudospectral,bugner2016solving,
bhattacharyya2021implementation,ruter2018hyperbolic,renkhoff2023adaptive}
is a numerical code that has been used successfully to study critical
collapse~\cite{hilditch2016pseudospectral,cors2023formulation,fernandez2022evolution,hilditch2017evolutions}.
It uses a pseudospectral collocation method for the spatial discretization
of the EFEs and an explicit fourth-order Runge-Kutta time stepping
algorithm.
\bamps{} utilizes distributed memory parallelization based on
the \textit{Message Passing Interface} (MPI) standard and, recently,
has been complemented with an adaptive mesh refinement (AMR) feature~\cite{renkhoff2023adaptive}.
However, in this work we do not make use of the AMR feature in order to facilitate the comparison.
This means that we are using the same static computational grid structure between
different simulations and we use the same polynomial resolution in all grid cells.
In \autoref{tab:grid-parameters} we summarize important
parameters of our computational setup.

The GHG formulation~\eqref{eq:ghg}-\eqref{eq:ghg3} of the EFEs
to evolve the gravitational field
and the first order reduction of the Klein-Gordon equation~\eqref{eq:kleingordon-reduction1}-\eqref{eq:kleingordon-reduction3}
to evolve the complex scalar field are implemented in \bamps{}.
The code employs radiation-controlling and constraint-preserving boundary conditions
as described in~\cite{hilditch2016pseudospectral,rinne2007testing,lindblom2006new}.
The scalar field uses a maximally dissipative boundary condition on the physical degrees
of freedom and constraint preserving boundary conditions for the reduction constraints~\cite{bhattacharyya2021implementation}.
We note that these boundary conditions were initially designed for real massless scalar-field
evolutions and were adapted for this study to also work for massless complex scalar fields.
In particular, they do not account for a scalar-field potential $V(|\phi|^2)$,
which is likely the cause for some of the artifacts we report below.
We leave it for future work to improve these conditions.
After extensive testing we found that the combination of the GHG formulation
with the constraint damping parameters as given in \autoref{Section:Theory}
together with the grid parameters reported in \autoref{tab:grid-parameters}
and the above boundary conditions allow us to perform long-time stable
and convergent evolutions of binary BS head-on collisions.

\section{Results}
\label{Section:Results}

\begin{table}[t]
 \centering
 \caption{
    Properties of the stationary, isolated and spherically symmetric mini BS solution
    that is used as the basis for all the binary BS initial data constructions
    presented in this work. The same configuration has been studied in~\cite{helfer2022malaise}.
    $A_{\textrm{ctr}}$ is the central scalar-field amplitude,
    $\omega$ is the stationary angular frequency of the harmonic time dependence of
    the scalar field, $M_{\textrm{BS}}$ is the ADM mass of the system and
    $r_{99}$ is the radius in which $99\%$ of the ADM mass are contained.
    We use $\MBBS = 2 M_{\textrm{BS}} = 0.79$ as an estimate for the
    initial ADM mass of the binary system.
  }
  \begin{tabularx}{\columnwidth}{XXXXX}
    \hline
      Model             & $A_{\textrm{ctr}}$  & $\omega$  & $M_{\textrm{BS}}$   & $r_{99}$\\
    \hline
      \texttt{mini1}    & 0.0124     & 0.97(1)   & 0.39(5)    & 22.(6)\\
    \hline
  \end{tabularx}
 \label{tab:bs-parameters}
\end{table}

We focus exclusively on head-on collisions of equal mass mini BSs.
Some of the key properties of the particular stationary and isolated BS solution we used as the basis
for the initial data construction are summarized in \autoref{tab:bs-parameters}.
In the head-on configurations we study we vary the initial boost parameter $v$
as well as the initial separation $d$ between the stars.
Similar configurations were discussed already in~\cite{helfer2022malaise}
and a slight variation thereof, with non-zero impact parameter, was investigated in~\cite{croft2023gravitational}.
All collisions for which we report results below culminate in
a single perturbed and non-spinning BS remnant with zero bulk motion with respect to the coordinate origin
and where the gravitational and scalar fields continue to oscillate.
The latter process causes a continued emission of GWs and scalar-field radiation, which lasts
for much longer than the head-on impact.
The GWs emitted due to the oscillating remnant are referred to as
the gravitational afterglow of BSs~\cite{croft2023gravitational}.
This afterglow radiation is characterized by an amplitude
that is comparable with the GW burst that is due to the initial head-on impact.
See \autoref{fig:cmon_psi4_longrun} below for an example of GW afterglow obtained
from a long-time evolution.

Besides the intended direct comparison of the initial data quality, we repeat some
of the experiments reported in~\cite{helfer2022malaise}
to perform calibration tests with our NR code \bamps{}, as it is the first time it
is used for BS evolutions.
Given that \bamps{} employs a PS method
for the spatial discretization of the fields, whereas~\cite{helfer2022malaise} used
the \textsc{Lean} code~\cite{sperhake2007binary}
and the \textsc{GRChombo} code~\cite{husa2008reducing,clough2015grchombo} which both work with
finite-differencing methods, the presented results also provide an (indirect) benchmark
between different numerical methods for the simulation of spacetimes with smooth matter fields.

To gauge differences in numerical evolutions for the comparison of
constraint-satisfying and violating initial data one can use various quantities.
Below we focus on a mixture of constraint monitors,
global (and conserved) physical quantities, as well as local field values
and gravitational waves.
Studying these quantities also allows us to assess the accuracy and reliability of the
PS scheme employed.
This is important to our work, because the equations of motion of
the matter (Klein-Gordon equation) can be rewritten to mimic a
\textit{balance law} (\ie{} a conservation law with an inhomogeneity)
for which a plethora of numerical methods have been developed in the
computational fluid dynamics literature and are often employed in the context
of general relativistic hydrodynamics simulations.
On the other hand the PS method developed in~\cite{hilditch2016pseudospectral}
was solely designed to conserve energy within a particular approximation
and so it is of interest to us to see how well the scheme can \textit{balance}
matter fields over long times.

During binary BS evolutions we compute:
\begin{enumerate}[i)]

\item The constraint monitor $\Cmon$ defined in~\cite{hilditch2016pseudospectral}.
It summarizes violations of the constraint subsystem of GHG,
among which are the Hamiltonian and momentum
constraints~\eqref{eq:hamilton-constraint}
and~\eqref{eq:momentum-constraint} as well as the
harmonic gauge constraint~\eqref{eq:harmonic-gauge-constraint}.
In the continuum limit $\Cmon = 0$ throughout the evolution.
In practice we observe non-zero values for $\Cmon$
and they serve as a proxy to gauge to which accuracy the EFEs can be solved.

\item The dynamical behavior of the stars and spacetime are monitored through
the maximum of the scalar-field amplitude $\Amax(t)$
and the value of the Ricci scalar at the coordinate origin $R(t,x=0)$.
These quantities together with $\Cmon$ are used to distinguish physical signatures
from numerical artifacts in our analysis.

\item The Noether charge associated with the global $U(1)$ symmetry of~\eqref{eq:action}
  is given by~\cite{liebling2023dynamical}
  \begin{align}
  N &= \int_V \dd x^3~\sqrt{-g} j^t \, ,
  \\
  j^a &= \frac{i}{2} g^{ab} \left(
    \phi^\ast \nabla_{b} \phi - \phi \nabla_{b} \phi^\ast
  \right) \, ,
  \label{eq:noether-current}
  \end{align}
where $j^a$ is the Noether current and integration is performed over the whole
computational domain $V$.
$N$ can be related to the total number of bosonic
particles~\cite{liebling2023dynamical,ruffini1969systems}.
This quantity is also conserved, provided no matter leaves the domain through a boundary,
and, thus, its time evolution allows to gauge the
accuracy of the evolution of the Klein Gordon equation.

\item The ADM mass of the spacetime~\cite{baumgarte2010numerical}
  \begin{align}
  \MADM
    &= \frac{1}{16\pi} \lim_{r\to\infty} \int_{\partial\Sigma_r} \dd S~ \mathcal{N}^k \gamma^{ij} (
      \partial_{j} g_{ik} - \partial_{k} g_{ij} ) \, ,
  \end{align}
where $\mathcal{N}^m$ is the outward pointing unit normal vector to
a 2-hypersurface $\partial\Sigma_r$ of a spatial slice $\Sigma$.
In theory, this quantity is conserved in time and
we monitor its temporal evolution to benchmark our results.
Note that the results of $\MADM$ we report below were obtained without taking the limit $r \to \infty$
and instead computed at a finite radius.
In this sense all references of $\MADM$ in the following refer to an approximation
of the ADM mass.
In fact, this approximation of $\MADM$ is also not necessarily conserved
and can show a decrease in time, in particular when matter leaves the computational domain.

\item A \textit{total mass} number~\cite{croft2023gravitational}
\begin{align}
  \Mtot &= \int_V \dd x^3 \sqrt{\gamma} \rho \, ,
\end{align}
where $\rho = T_{\mu\nu} n^\mu n^\nu$ is the local energy density and integration is
performed over the whole computational domain $V$.
This is a coordinate dependent quantity and it is not necessarily conserved.

\item Gravitational radiation represented through the curvature pseudo-scalar field $\Psi_4$.
This (or rather integrals thereof) is the only accessible observable
with which binary BS encounters could be experimentally detected.
In particular, we only focus on the dominant $l,m=2,0$ mode.
We leave out a discussion of the GW strain $h$ due to unacceptably large uncertainties
introduced in the reconstruction procedure, see \aref{Appendix:gw-analysis}.

\item The radiated GW energy $E$ associated with $\PsiFT$ as recorded by
an asymptotic observer over a fixed time interval.
This quantity was also studied in~\cite{helfer2022malaise} to analyze the
behavior of the long-lived oscillating remnant.
The error bars we provide account for
errors due to finite resolution, errors in the extrapolation to null infinity
and errors in the reconstruction of $E$ from $\PsiFT$.
For details on this analysis see \aref{Appendix:gw-analysis}.

\item The detector-noise-weighted Wiener product $W(\PsiFT^{(1)}, \PsiFT^{(2)})$ between two GW signals.
This quantity gives a measure of how \textit{similar} two GWs are, while accounting for
detector sensitivity.
To evaluate $W$ one has to assume a value for scalar-field mass $\mu$
in order to convert the results to SI units.
Because we use a noise-sensitivity curve of Advanced LIGO~\cite{ligopsd} to weight the Wiener product,
we fix $\mu = 1\times 10^{-11}~\text{eV}$ so that the frequency of the dominant component
of the PSD of $\PsiFT$ falls into the most sensitive region of the detector, which is at $\mathcal{O}(100~\text{Hz})$~\cite{aligo2020}.
For details see \aref{Appendix:gw-wiener-product}.

\end{enumerate}

\subsection{Constraint violations}

\begin{figure*}[t]
  \centering
  \includegraphics[width=\textwidth]{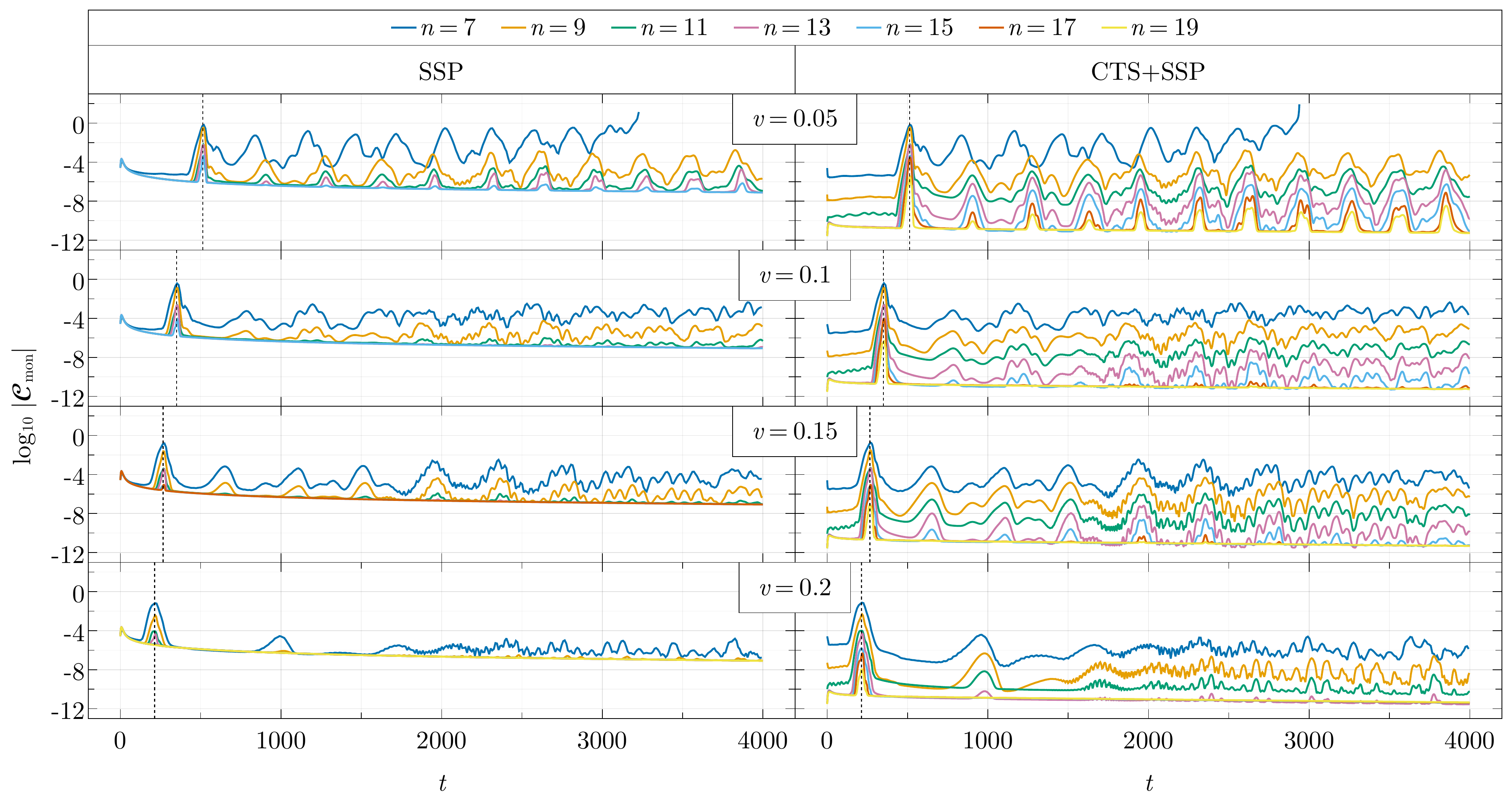}
  \caption{
  Side by side comparison of the time evolution of $\Cmon$
  for simulations based on SSP data (left column) and CTS+SSP data (right column)
  for fixed initial separation $d = 80$, varying resolutions and
  different values of the boost parameter $v=$ 0.05, 0.1, 0.15, 0.2 (from top to bottom).
  Vertical dashed lines indicate time of merger as determined by the time of the maximum value of $\Amax(t)$
  from the highest resolution.
  }
  \label{fig:ana_cmon_compare_resolutions_naive}
  \vspace{3mm}
  \includegraphics[width=\textwidth]{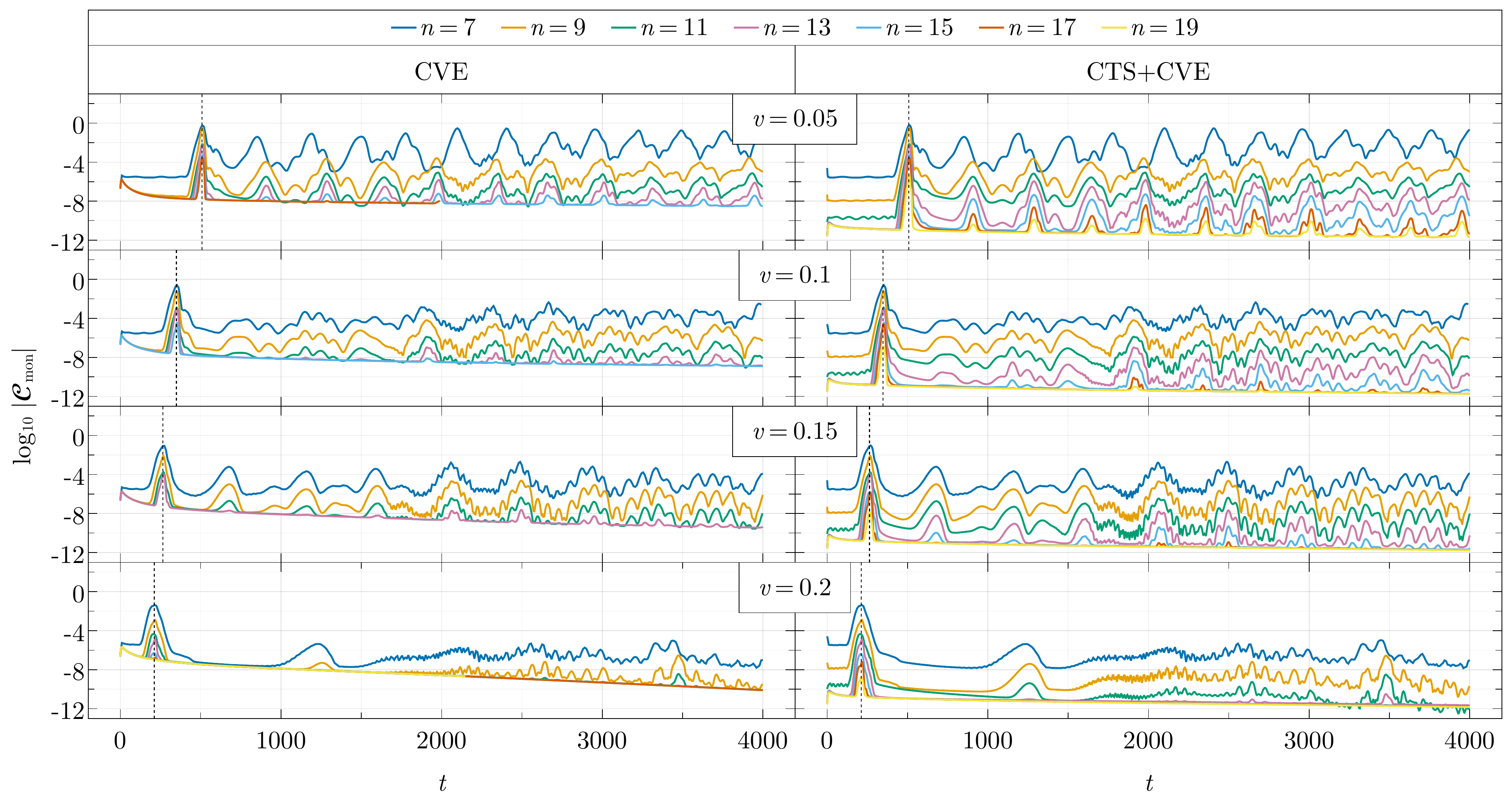}
  \caption{
  Side by side comparison of the time evolution of $\Cmon$
  for simulations based on CVE data (left column) and CTS+CVE data (right column)
  for fixed initial separation $d = 80$, varying resolutions and
  different values of the boost parameter $v=$ 0.05, 0.1, 0.15, 0.2 (from top to bottom).
  Vertical dashed lines indicate time of merger as determined by the time of the maximum value of $\Amax(t)$
  from the highest resolution.
  }
  \label{fig:ana_cmon_compare_resolutions_helfer}
\end{figure*}

\begin{figure*}[t]
  \begin{subfigure}[b]{0.49\textwidth}
    \includegraphics[width=\linewidth]{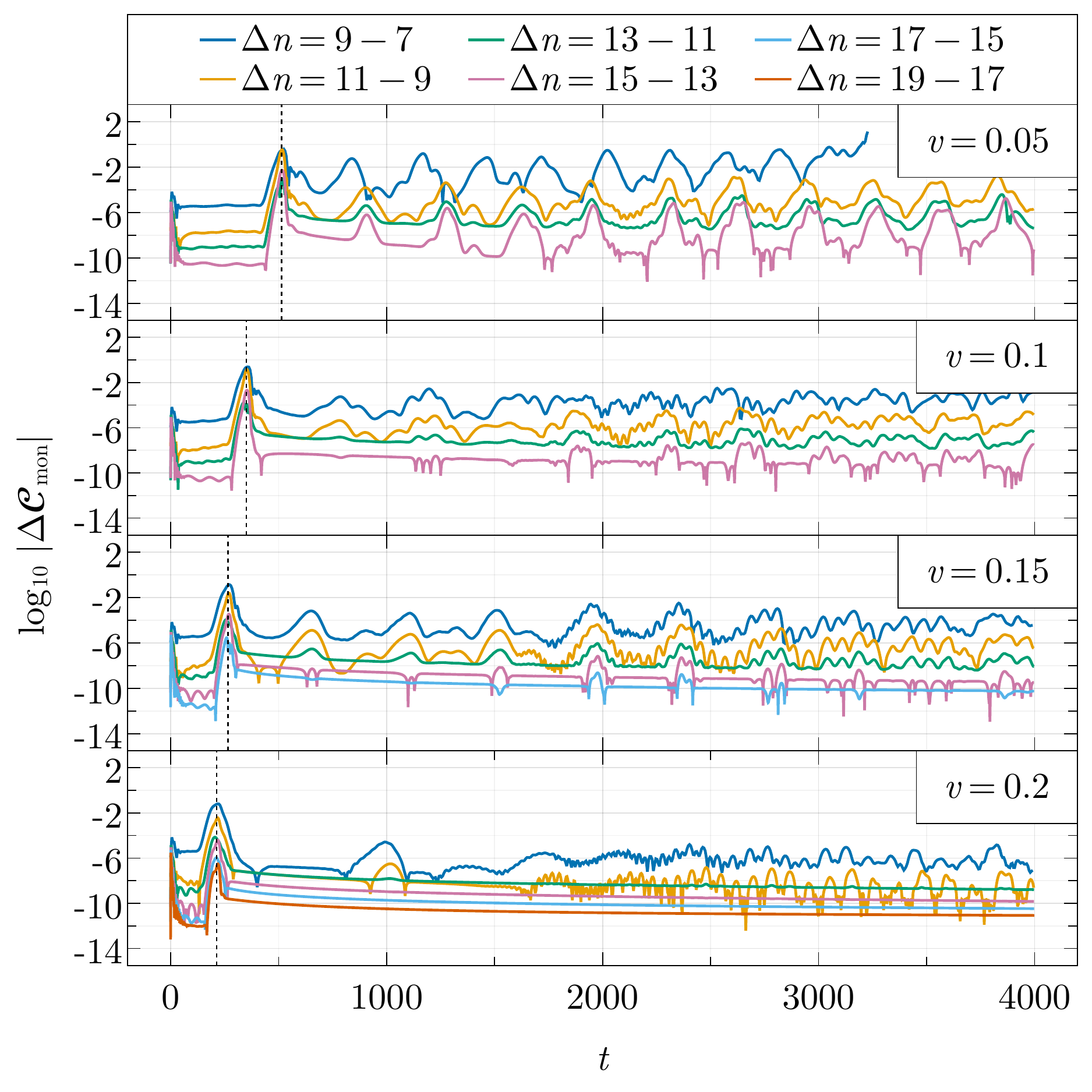}
  \end{subfigure}
  \hfill
  \begin{subfigure}[b]{0.49\textwidth}
    \includegraphics[width=\linewidth]{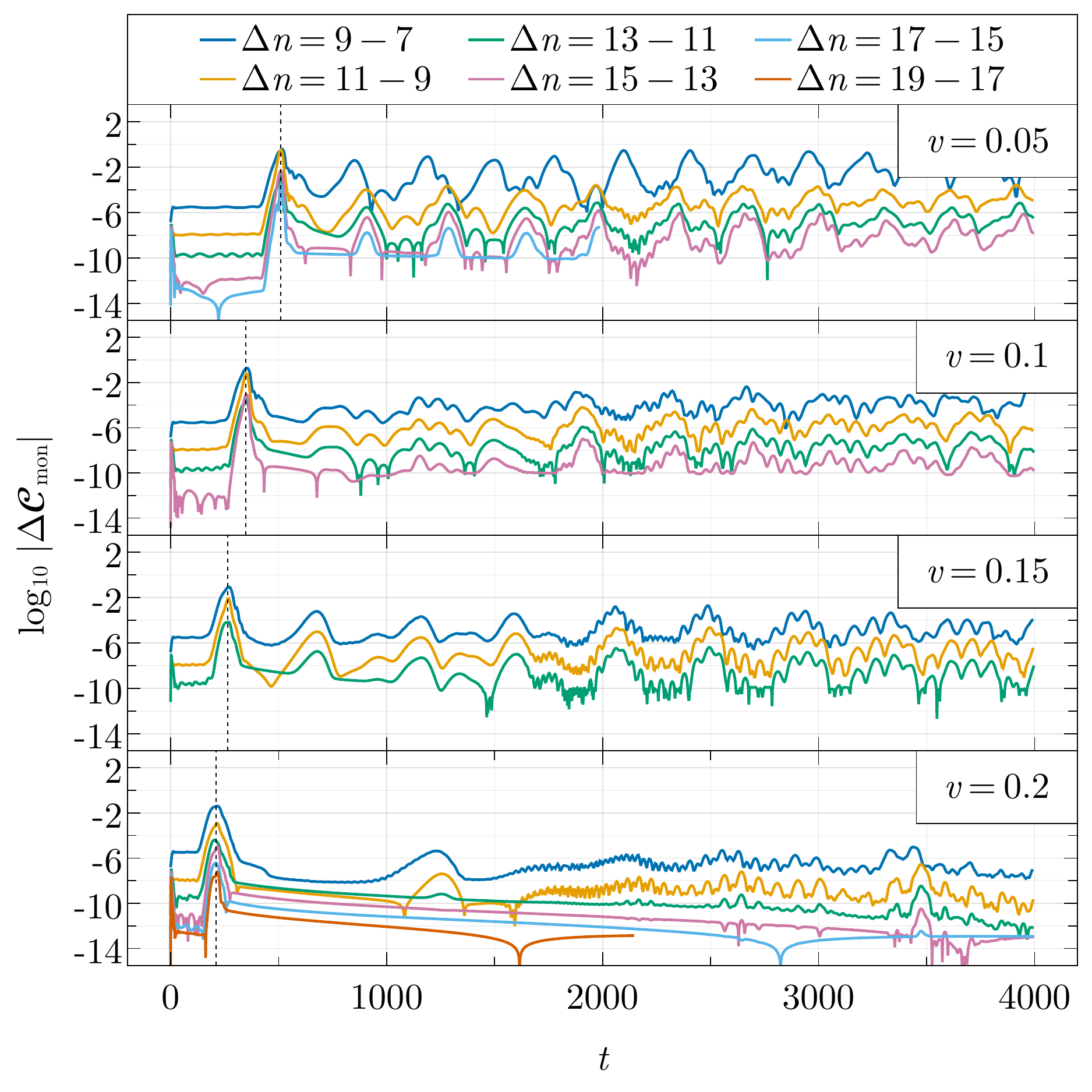}
  \end{subfigure}
  \caption{
    Self-convergence test of the constraint monitor $\Cmon$.
    \textit{Left} Results obtained from evolutions of SSP data.
    \textit{Right} Results obtained from evolutions of CVE data.
    The initial separation is fixed to $d = 80$ and initial boost values are $v=$ 0.05, 0.1, 0.15, 0.2.
    Vertical dashed lines indicate time of merger as determined by the time of the maximum value of $\Amax(t)$
    from the highest resolution.
  }
  \label{fig:diff_anacmon_convergence_study}
\end{figure*}

\begin{figure*}[t]
  \begin{subfigure}[b]{0.49\textwidth}
    \includegraphics[width=\linewidth]{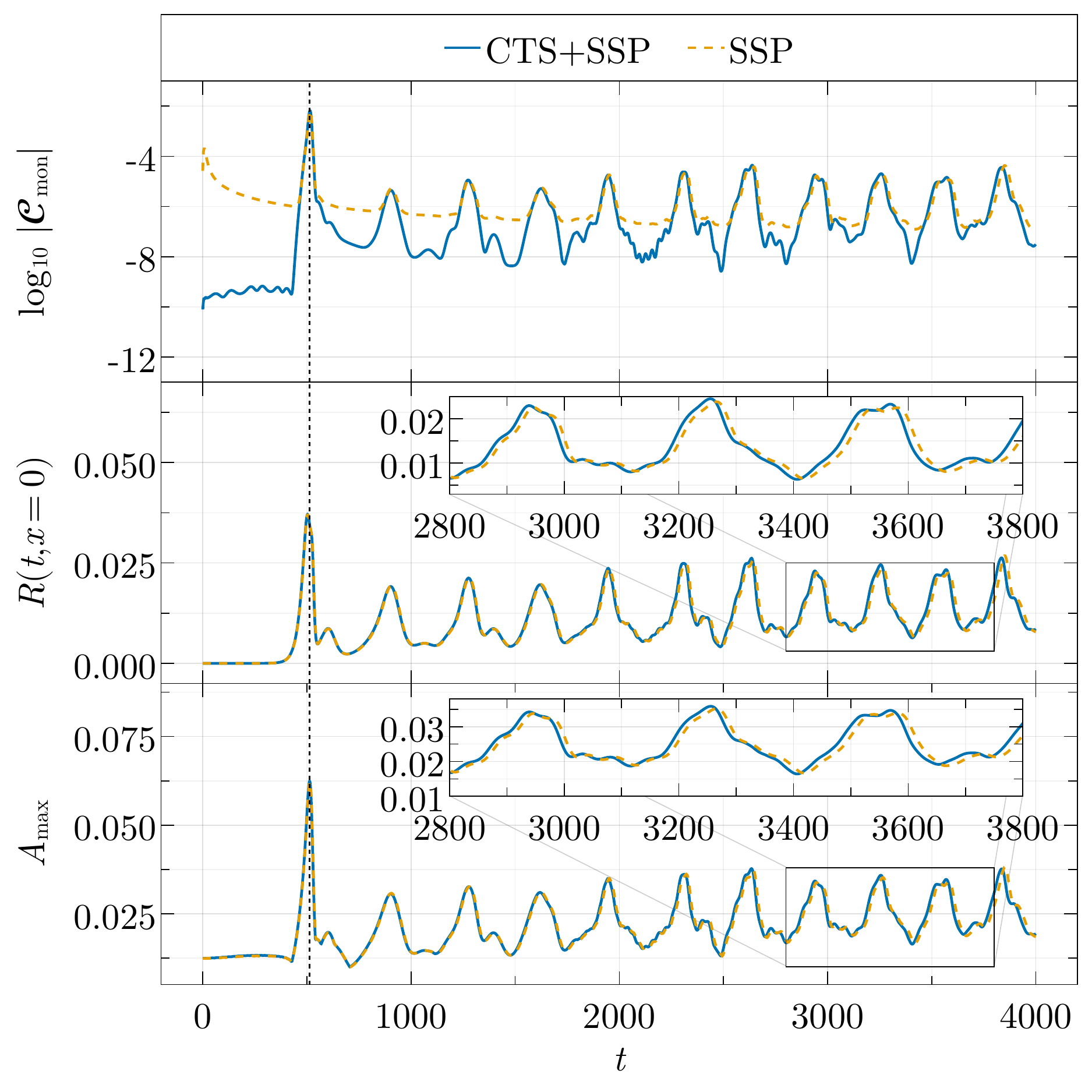}
  \end{subfigure}
  \hfill
  \begin{subfigure}[b]{0.49\textwidth}
    \includegraphics[width=\linewidth]{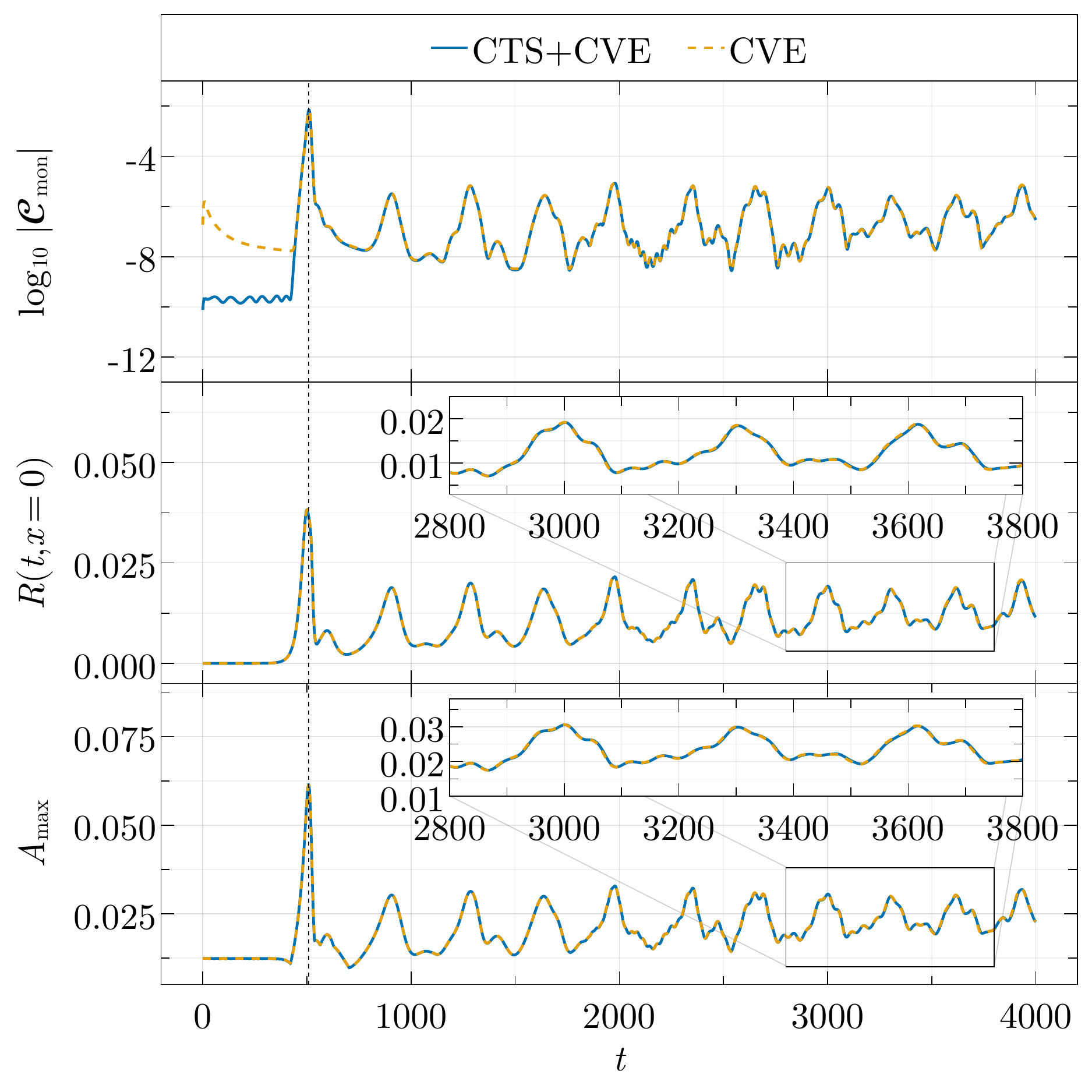}
  \end{subfigure}
  \caption{
    Direct comparison of evolution quantities
    $\Cmon$ (top), $R(t,x=0)$ (middle) and $\Amax(t)$ (bottom)
    obtained from binary BS evolutions of superposed initial data (dashed)
    \vs{} constraint-solved data (solid).
    The stars were prepared with boost parameter $v = 0.05$ and initial distance $d = 80$
    and polynomial resolution was $n = 11$.
    \textit{Left} Comparison between SSP data \vs{} and CTS+SSP data.
    \textit{Right} Comparison between CVE \vs{} and CTS+CVE data.
    The vertical dashed lines indicate time of merger as determined by the time of the maximum value of $\Amax(t)$.
  }
  \label{fig:cmon_rsc_amppsi_comparison}
\end{figure*}

\begin{figure}[t]
  \includegraphics[width=0.99\columnwidth]{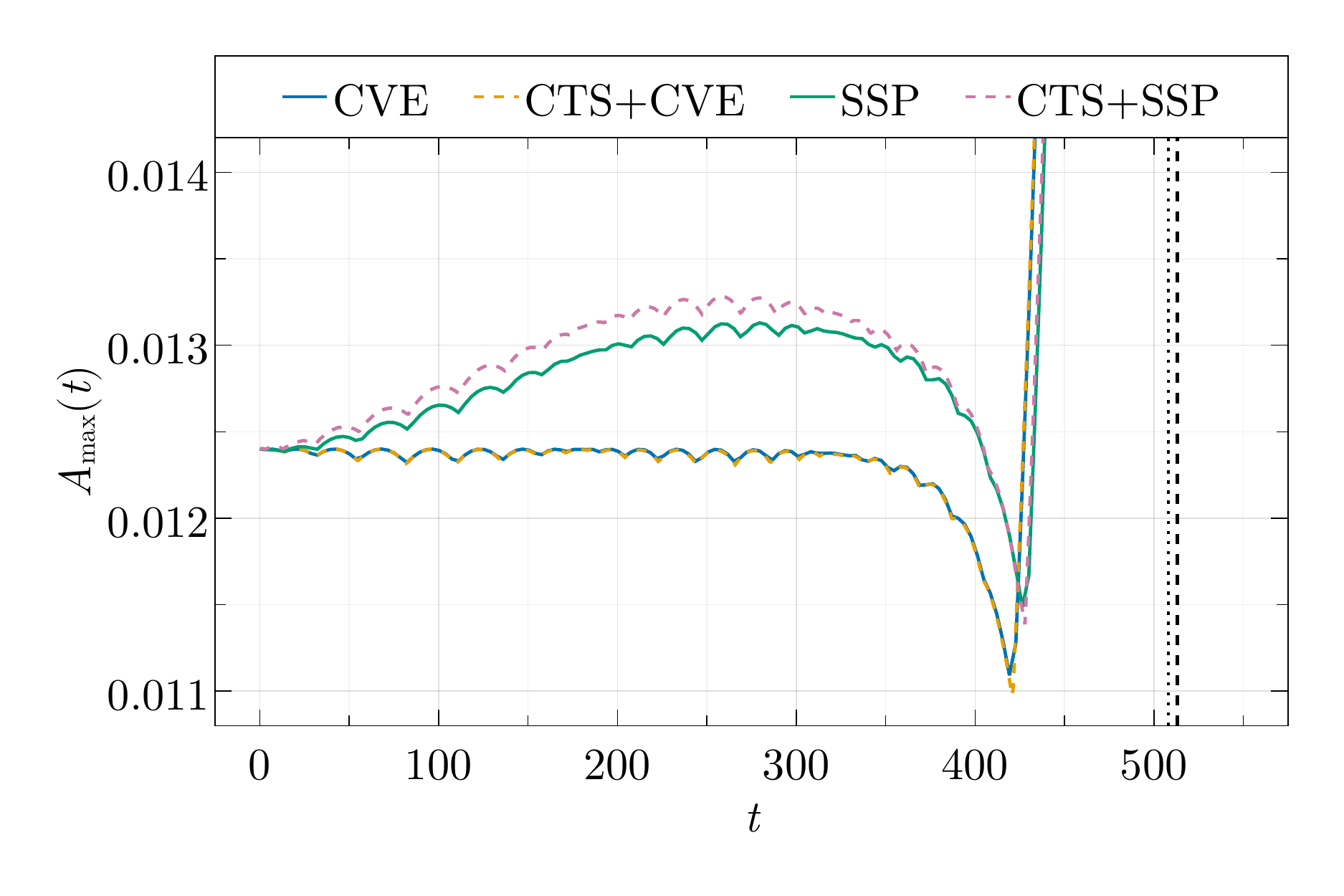}
  \caption{
    Direct comparison of $\Amax(t)$ obtained from binary BS evolutions
    that used constraint-violating SSP and CVE data (solid) and
    constraint-satisfying CTS+SSP and CTS+CVE data (dashed).
    The stars were prepared with boost parameter $v = 0.05$, initial distance $d = 80$
    and polynomial resolution was $n = 11$.
    This is a zoomed in plot of the $\Amax(t)$ results presented in \autoref{fig:cmon_rsc_amppsi_comparison}.
    The vertical dashed and dotted lines indicate time of merger as determined
    by the time of the maximum value of $\Amax(t)$ from the CTS+SSP and CTS+CVE data, respectively.
  }
  \label{fig:amax_infall_v_0.05_n_11}
\end{figure}

A comparison of the convergence behavior of $\Cmon$ obtained from runs
that used SSP data \vs{} CTS+SSP data with initial separation $d = 80$,
boost parameters $v = 0.05,0.1,0.15,0.2$ and varying polynomial resolutions $n$
is provided in \autoref{fig:ana_cmon_compare_resolutions_naive}.

First, we compare the values of $\Cmon$ at $t = 0$ between the constraint-violating
SSP data (left columns) and constraint-satisfying CTS+SSP data (right columns).
The figure shows that all SSP evolutions start from an initial violation
$\Cmon \approx 10^{-4}$, which is independent of the resolution and initial boost.
On the other hand, the CTS+SSP data starts off at $\Cmon \approx 10^{-5}$ and
decreases with $n$ and independent of $v$ to $\Cmon < 10^{-10}$.
Because we solve the CTS equations for each resolution $n$ separately, instead of
solving them once for a high resolution and then using interpolation to obtain data
on a lower resolution, this behavior of $\Cmon$ demonstrates that our CTS solver
is capable of removing excess constraint violations with increased resolution.

Focusing on the time evolution of $\Cmon$ of CTS+SSP data,
one observes a clear convergence pattern in resolution.
We want to emphasize the rate by which $\Cmon$ decreases
by pointing to the exponential improvement of $\Cmon$ from $10^{0}$ down to $10^{-12}$
while the polynomial resolution $n$ increases linearly in steps of two from 7 to 19
(see color coding in legend).
For boost values $v =$ 0.1, 0.15 and 0.2 one observes that with resolutions $n \geq 15$
the violations are bounded by $\Cmon \lesssim 10^{-10}$ when $0 \leq t \leq 4000$,
except around the merger which occurs at
$t \approx 350$, 260 and 210, respectively.
For the runs using $v = 0.05$ one would need to increase the resolution beyond $n = 19$ to
achieve the same level of violations in the afterglow signal, which is computationally
well within reach even without AMR.
Comparing now with the violations from evolutions of SSP data
one can also see that $\Cmon$ decreases for all simulations independent of $v$.
However, the results for $v=$ 0.05, 0.1 and 0.15 show that this trend eventually halts when $n \gtrsim 13$
and for $v=0.2$ no improvement occurs beyond $n = 11$ in the afterglow.
From this we conclude that evolutions of non-constraint-solved SSP data bear
a residual constraint violation $\Cmon \gtrsim 10^{-8}$ when $0 \leq t \leq 4000$
for this configuration in our code.
Note that this result does not imply that these evolutions are not convergent,
because a numerical value of order $\mathcal{O}(10^{-8})$ is usually
not regarded as \textit{numerically zero}.
Instead one must treat this data series as a series that approaches a non-zero
residual value and conduct a self-convergence test, which is done further below.

Studying now the behavior of $\Cmon$ around merger
one observes that the constraint monitor continues to improve with increasing $n$,
even when $\Cmon$ was already saturated away from merger.
This behavior is independent of the boost parameter $v$ and occurs
for SSP and CTS+SSP data.
The reason for this is that the merger phase typically involves higher field
amplitudes and gradients compared to less extreme field configurations before and after merger,
which translates into a need for locally higher polynomial approximations
to resolve the solutions accurately,
due to aliasing effects in the PS expansion being amplified by the nonlinearity of
the EFEs.
As a side note we mention that the use of AMR
could potentially bring down these constraint violations around merger to the same level away from the merger,
while at the same time improve computational efficiency by distributing the available
resources where numerical resolution is needed. We leave this test to future work as
it will certainly become of relevance for inspiral simulations.

\autoref{fig:ana_cmon_compare_resolutions_helfer} shows the same comparison of $\Cmon$,
but for evolutions that started with constraint-violating CVE data \vs{}
constraint-satisfying CTS+CVE data and the same initial distance and boost parameters.
The overall behavior of the results is similar to the one from
\autoref{fig:ana_cmon_compare_resolutions_naive}:
at $t = 0$ the CVE data comes with an initial violation $\Cmon \approx 10^{-6}$
that is independent of $n$ and $v$, whereas the CTS+CVE data starts at $\Cmon \approx 10^{-5}$
and continuously decreases with increasing resolution to $\Cmon < 10^{-10}$.
The time evolution of CTS+CVE data also displays a convergence pattern
with exponential decrease when increasing $n$, and the constraint violations
are reduced to $\Cmon < 10^{-10}$ when $0 \leq t \leq 4000$, except around merger.
Also similar is the behavior of the results from constraint-violating CVE
initial data where it is evident that increasing $n$ decreases $\Cmon$.
The plot also shows that the constraint monitor saturates for $\Cmon > 10^{-10}$ when $v = $ 0.05, 0.1 and 0.15.
When $v = 0.15$ or 0.2 we also observe that the lower bound on $\Cmon$
additionally decreases over time and approaches a value of $\approx 10^{-10}$ at $t = 4000$.
As for the merger phase, increasing $n$ also reduces $\Cmon$ for both CVE and CTS+CVE data.

In \autoref{fig:diff_anacmon_convergence_study} a self-convergence test
of $\Cmon$ using the constraint-violating data presented in
\autoref{fig:ana_cmon_compare_resolutions_naive} (left columns) and
\autoref{fig:ana_cmon_compare_resolutions_helfer} (left columns)
is depicted.
With such a test one can study the convergence of a series without knowing
the exact result the series is converging to.
This method is applied to $\Cmon$ here, because for constraint-violating data
$\Cmon$ does not approach zero, but it attains
limiting values $\Cmon \gtrsim 10^{-8}$ and $\Cmon \gtrsim 10^{-10}$ for SSP and CVE data,
respectively.
\autoref{fig:diff_anacmon_convergence_study} demonstrates that the differences
of $\Cmon$ between consecutive resolutions
decrease exponentially with increasing resolution, thus, confirming the
claim made earlier that $\Cmon$ is also convergent for evolutions of initially-constraint-violating data.

A direct comparison of the analysis quantities $\Cmon(t)$, $R(t,x=0)$ and $\Amax(t)$
obtained from evolutions starting from constraint-violating and
constraint-solved initial data,
based on a configuration with initial separation $d = 80$,
initial boost $v = 0.1$ and polynomial resolution $n=11$,
is provided in \autoref{fig:cmon_rsc_amppsi_comparison}.
Focusing on the results obtained with the SSP construction method (left column), it is evident
that $\Cmon$ is always smaller for the evolution of CTS+SSP data than the
evolution of SSP data when $0 \leq t \leq 4000$.
The difference in $\Cmon$ between these evolutions gradually decreases over time
and there is considerable overlap in the afterglow phase of the evolution.
However, this overlap does not continue to hold with increasing polynomial resolution $n$,
because $\Cmon$ for the runs with SSP data eventually levels off at a non-zero value,
as demonstrated previously.
Comparing vertically $\Cmon(t)$ (top, left) \vs{} $R(t,x=0)$ (middle, left) \vs{} $\Amax(t)$ (bottom, left)
one can see that the times at which the constraint monitors overlap
correspond to times where $R(t,x=0)$ and $\Amax(t)$ each show local maxima.
This allows to conclude that the increase in $\Cmon$ around merger, as well as the
oscillations in the afterglow, are caused by physical processes.
Studying the time evolutions of $R(t,x=0)$ and $\Amax(t)$ closer
one can observe good agreement between SSP and CTS+SSP until
$t \approx 2800$, after which these quantities start to dephase (see insets).
This dephasing persists when increasing the resolution.
Not visible in this plot is that $\Amax(t)$ also shows deviations during the infall
phase when $0 \leq t \leq 500$, because they are about a factor of 10 smaller than the
differences that appear at late times and we revisit this further below.

The comparison of $\Cmon$ from runs with constraint-violating and constraint-solved initial data
based on the CVE construction is displayed in \autoref{fig:cmon_rsc_amppsi_comparison} (right column).
Similar to the case of evolutions of SSP data, one observes that, prior to merger,
$\Cmon$ is always smaller for the evolutions using CTS+CVE data compared to
evolutions of CVE data. We see that at this resolution the evolution of $\Cmon$
after the merger is independent of the initial value of $\Cmon$ and, therefore, it is only dominated
by violations that occur during the merger.
Also similar is the temporal alignment in the extrema between $\Cmon(t)$, $R(t,x=0)$ and $\Amax(t)$.
The evolutions of $R(t,x=0)$ and $\Amax(t)$ also show that no dephasing occurs for these
quantities for late times at this resolution (as well as resolutions $n=9,13,15$, but not shown)
and, thus, one can conclude that the reduction in initial constraints
did not have a noticeable influence on these observables.
Comparing the late time behavior of $R(t,x=0)$ and $\Amax(t)$ between SSP data (left column) and CVE data (right column)
one can see from the insets that the amplitudes of these quantities show a difference,
in particular the strength of the oscillations is reduced for the CVE data.
This observation suggests that the effect of constraint-solving data has less impact
on physical observables than the way in which the superpositions are constructed.

\autoref{fig:amax_infall_v_0.05_n_11} shows another comparison of the evolution of $\Amax(t)$
as presented in \autoref{fig:cmon_rsc_amppsi_comparison}, but now with an emphasize
on the infall phase of the head-on collision where $0 \leq t \leq 500$.
This plot is added to make a connection with~\cite{helfer2022malaise} where
the authors found that, in head-on collisions of solitonic BSs,
the use of SSP data causes premature BH collapse, \ie{} an apparent horizon is detected
before the center of the stars even meet.
This behavior is preceded by a growth of the central scalar field amplitude before the collapse.
In their work, the evolutions of CVE initial data with the same solitonic BSs did not show this phenomenon,
and instead the central scalar field amplitude remained to good approximation constant during
all of the infall phase, only changing after the star's centers merged and before the
first apparent horizon was detected.
In \autoref{fig:amax_infall_v_0.05_n_11} one can see that also for a head-on collision
of two mini BSs, which is based on SSP initial data,
$\Amax(t)$ shows a growth during the infall phase.
Furthermore, it is evident that $\Amax(t)$ grows strongly even for CTS+SSP data.
In contrast to this, $\Amax(t)$ obtained from the evolution of CVE and CTS+CVE initial data
agrees between each other, it remains roughly constant during the infall phase.
Evenutally, at $t \approx 300$ one can see that $\Amax(t)$ drops down
and then increases again right before the time of merger at $t \approx 510$.
Some oscillations of $\Amax(t)$ are also visible, but they are likely due
to the fact that neither initial data construction accounts for any quasi-equilibrium conditions.
Although the mini BS evolutions considered in this work do not undergo BH formation,
the plot shows a similar qualitative behavior
of the scalar field amplitude as reported in~\cite{helfer2022malaise},
which is also independent of resolution (not shown).
Since we are interested in studying how two initially non-interacting stars fall in on each other,
we expect on physical grounds that the scalar field amplitude is not drastically altered
before the stars' centers get close to each other.
For the case at hand, the isolated stars have a radius of $r_{99} \approx 22$ and are
initially well separated by a distance $d = 80$.
Because $\Amax(t)$ is less for the CVE and CTS+CVE data than
for SSP and CTS+SSP data, we conclude that this result favors the use of the two former
initial data sets for mini BS head-on collisions,
assuming no additional considerations regarding quasi-equilibrium conditions are involved.

\begin{figure}[t]
  \begin{subfigure}[b]{0.49\textwidth}
    \includegraphics[width=\linewidth]{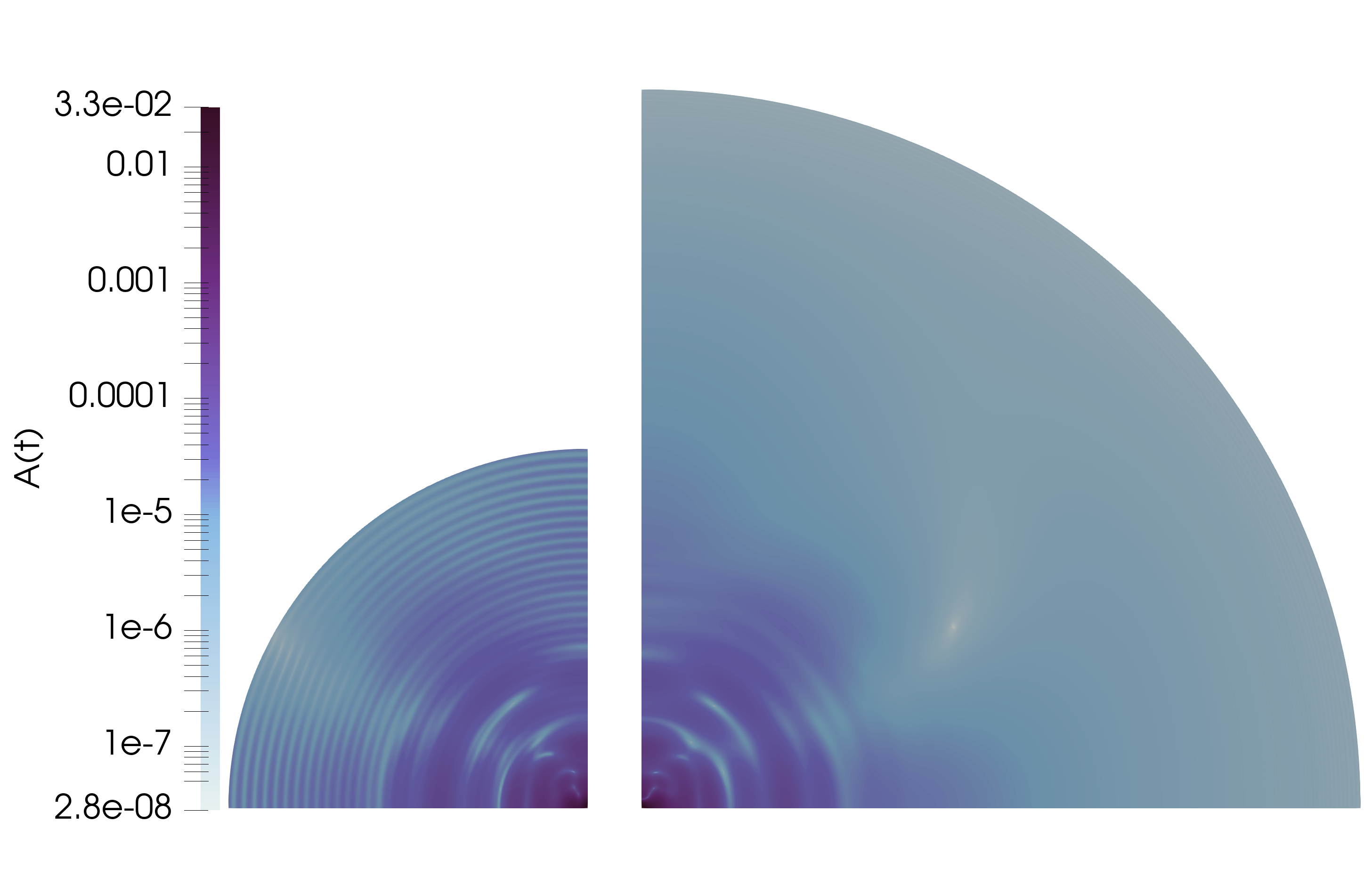}
  \end{subfigure}
  \begin{subfigure}[b]{0.49\textwidth}
    \includegraphics[width=\linewidth]{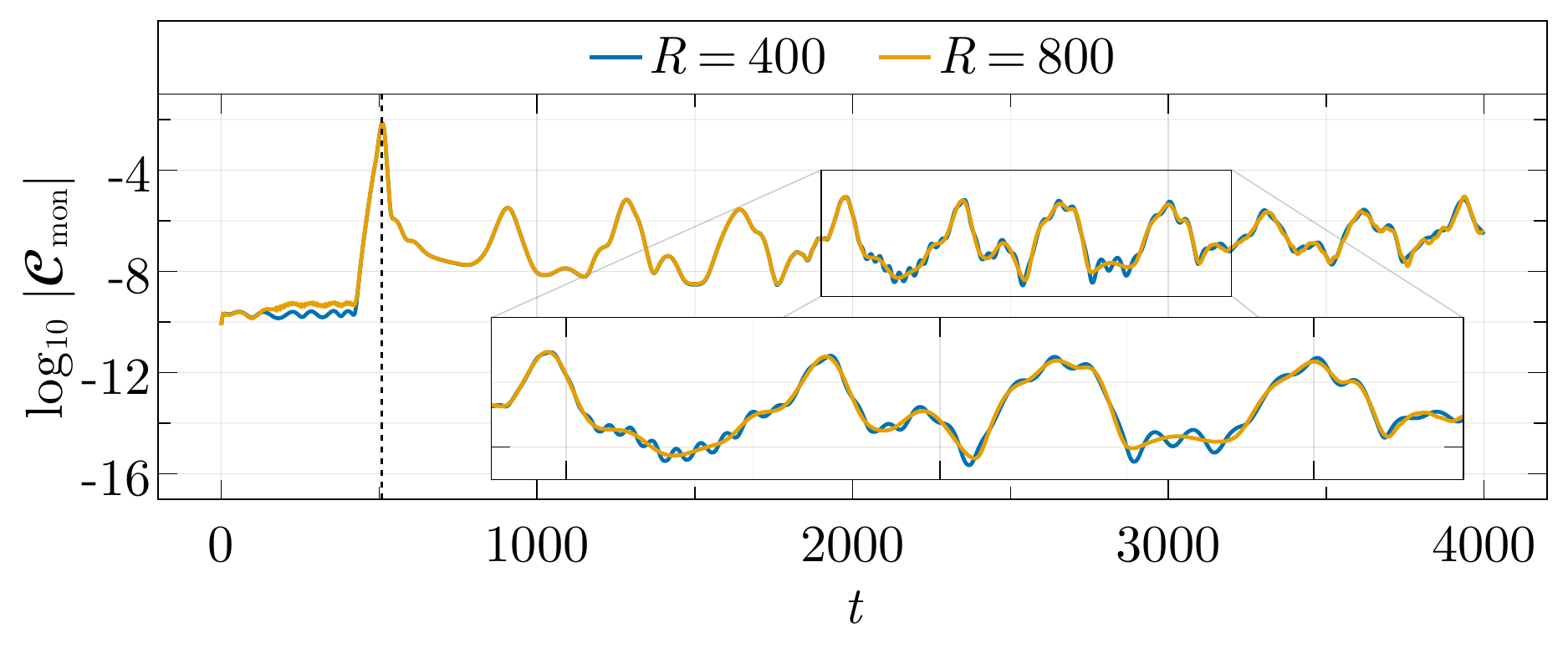}
  \end{subfigure}
  \caption{
    Results from evolution of CTS+CVE data with initial distance $d = 80$
    and boost parameter $v = 0.05$ for polynomial resolution $p=11$
    and two differently sized computational domains.
    \textit{Top} Comparison of the spatial distribution of scalar-field amplitude $A(t \approx 2000,x)$,
    domain radius $R = 400$ (left) and domain radius $R = 800$ (right).
    \textit{Bottom} Comparison of the evolution of $\Cmon$ for the differently sized
    computational domains.
    The vertical dashed line indicates the time of merger
    as determined by the time of the maximum value of $\Amax(t)$.
  }
  \label{fig:amppsi_2d}
\end{figure}

During testing we found that some observables can be polluted with artificial high frequency noise
for late times. As an example see $\Cmon$ in \autoref{fig:cmon_rsc_amppsi_comparison}
for SSP data (top, left) and CVE (top, right) which shows this noise starting to appear at
$t \approx 2000$.
These artifacts are not physical, but depend on parameters like the initial separation and boost
of the stars, and numerical resolution and are caused
by the scalar field interacting with the outer boundary conditions, which happens because
part of the scalar field is ejected from the central region during and after merger.
This hypothesis is confirmed by redoing these simulations with a grid setup where the
outer domain boundary is moved from radius $R=400$ to $R=800$.
Results of this test are displayed in \autoref{fig:amppsi_2d}, which
shows the spatial distribution of the scalar-field amplitude $A(t,x)$ (top)
extracted at coordinate time $t \approx 2000$ for the two differently sized domains.
It is evident that $A(t\approx2000,x)$ in the smaller domain displays noticeable
radial oscillations, whereas they are absent from the scalar-field profile evaluated
at the same coordinate time in the larger domain, which
reveals these oscillations as artifacts.
The comparison of the time evolutions of $\Cmon$ (bottom) extracted from these two simulations
shows that around $t \approx 2000$ the data is free of noise when the boundary is
placed at $R = 800$.
We note that this strategy of moving the boundary out further and further is in general
not a reliable solution when interested in artifact-free long-time evolutions,
because the associated computational costs eventually start to become prohibitive.
Although it might be possible to mitigate the computational costs by using AMR
with properly tuned $hp$ refinement,
a more sustainable solution would be to investigate how our boundary conditions need to be adjusted
to work with massive complex scalar fields. We leave this task to future work.
In the following discussion of global quantities and gravitational waves we show results that were
obtained with $R = 800$.

\subsection{Global quantities}

\begin{figure}[t]
  \includegraphics[width=0.99\columnwidth]{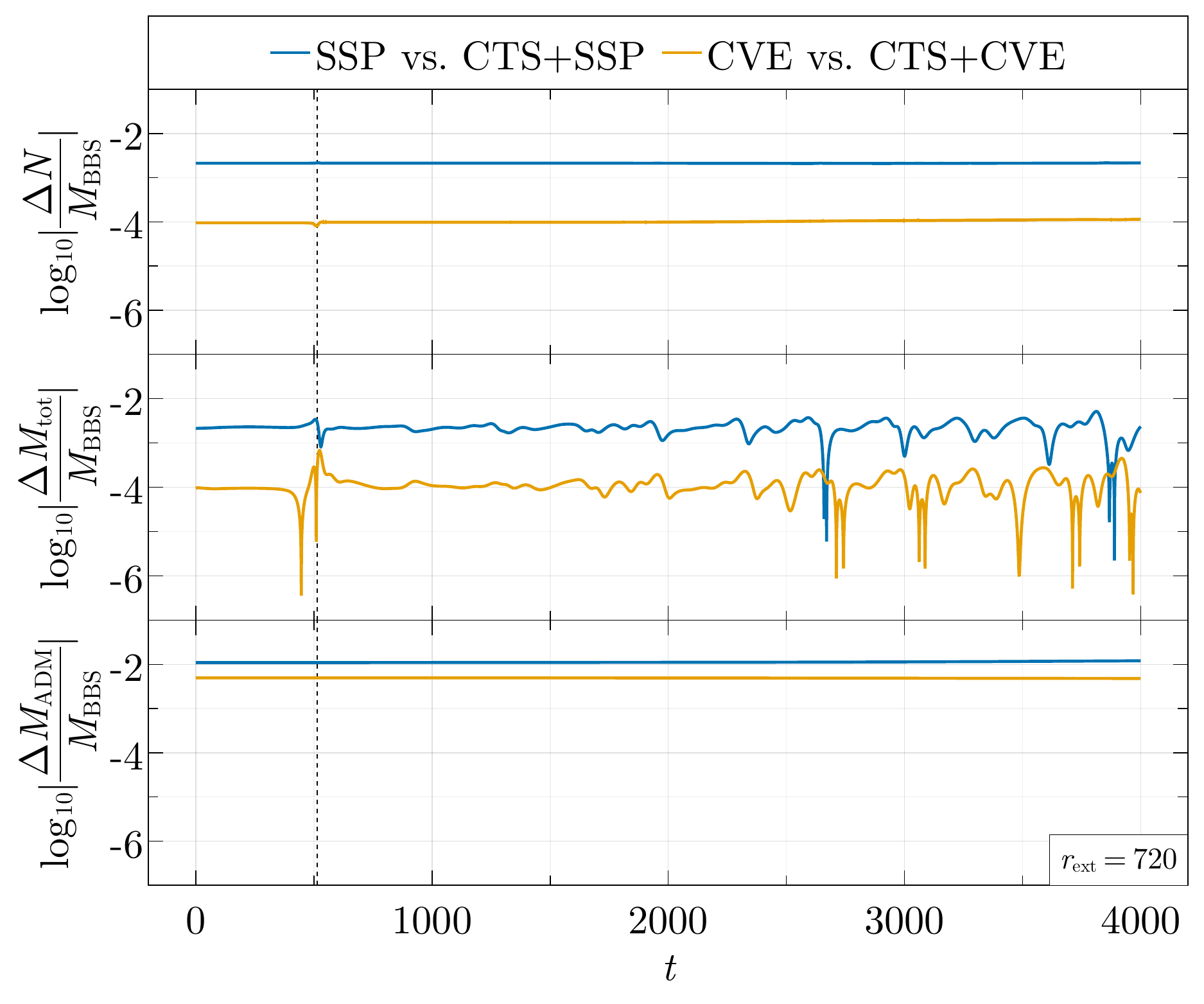}
  \caption{
    Relative differences of global quantities
    $N$ (top), $\Mtot$ (middle), $\MADM$ (bottom)
    between results from evolutions of SSP data \vs{} CTS+SSP and
    evolutions of CVE data \vs{} CTS+CVE data.
    The stars were prepared with initial boost parameter $v = 0.05$ and
    initial separation $d = 80$ and the differences were computed
    at polynomial resolution $n=11$.
    $\MADM$ was extracted at a sphere with coordinate radius $\Rext = 720$.
    The vertical dashed line indicates the time of merger
    as determined by the time of the maximum value of $\Amax(t)$.
  }
  \label{fig:comparison_N_MADM_Mtot}
\end{figure}

In \autoref{fig:comparison_N_MADM_Mtot} we continue the comparison started
in \autoref{fig:cmon_rsc_amppsi_comparison} by studying how relative differences
in global quantities develop for evolutions that used constraint-violating \vs{}
constraint-satisfying initial data with initial separation $d = 80$,
boost $v = 0.05$ and resolution $n=11$.
It is evident from the plot that the relative differences in
Noether charge $\Delta N$, ADM mass $\Delta \MADM$ and
total mass $\Delta \Mtot$ are constant in time to a good approximation.
The differences between CVE \vs{} CTS+CVE data evolutions
for $\Delta N$ and $\Delta \Mtot$ are of the order $0.01\%~\MBBS$,
whereas the differences between SSP \vs{} CTS+SSP evolutions
are below $0.1\%~\MBBS$.
The differences in $\Delta \MADM$ are of order $0.1\%~\MBBS$ and $1\%~\MBBS$ for
CVE \vs{} CTS+CVE and SSP \vs{} CTS+SSP, respectively.
We want to emphasize that the utilized resolution $n = 11$ is rather coarse
and computationally inexpensive, but the observed differences are already
very small for both data sets.

\begin{figure}[t]
  \includegraphics[width=0.99\columnwidth]{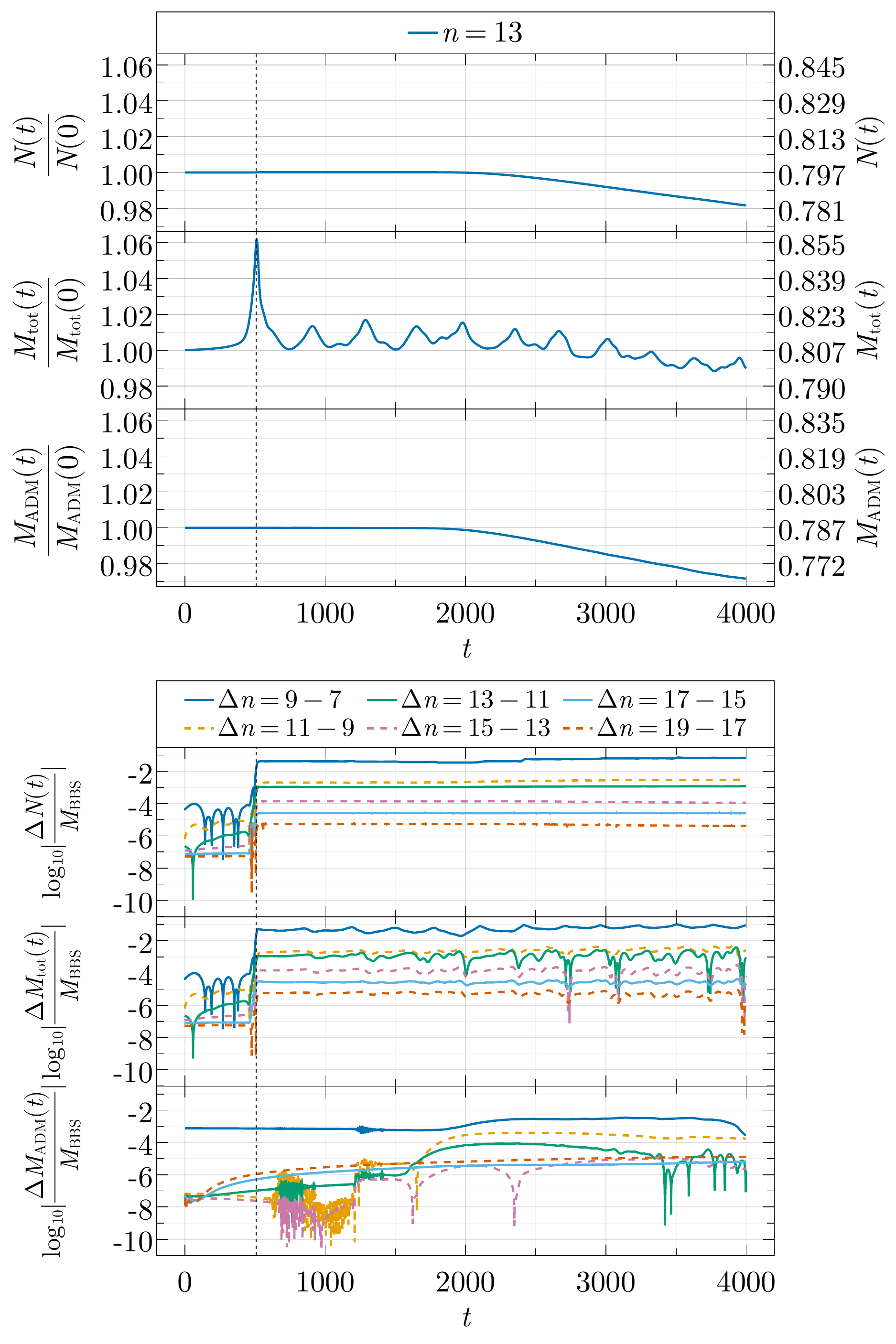}
  \caption{
    \textit{Top} Time evolution of the global quantities $N$, $\Mtot$ and $\MADM$
    obtained from a simulation with resolution $n=13$.
    \textit{Bottom} Self-convergence test of those variables.
    Results are from simulations of CTS+CVE data with initial separation $d = 80$
    and initial boost parameter $v = 0.05$.
    $\MADM$ was extracted at a sphere with coordinate radius $R = 720$.
    The vertical dashed lines indicate the time of merger
    as determined by the time of the maximum value of $\Amax(t)$ from the highest resolution.
  }
  \label{fig:convergence_totalM_noetherQ}
\end{figure}

\begin{figure*}[t]
  \centering
  \includegraphics[width=\textwidth]{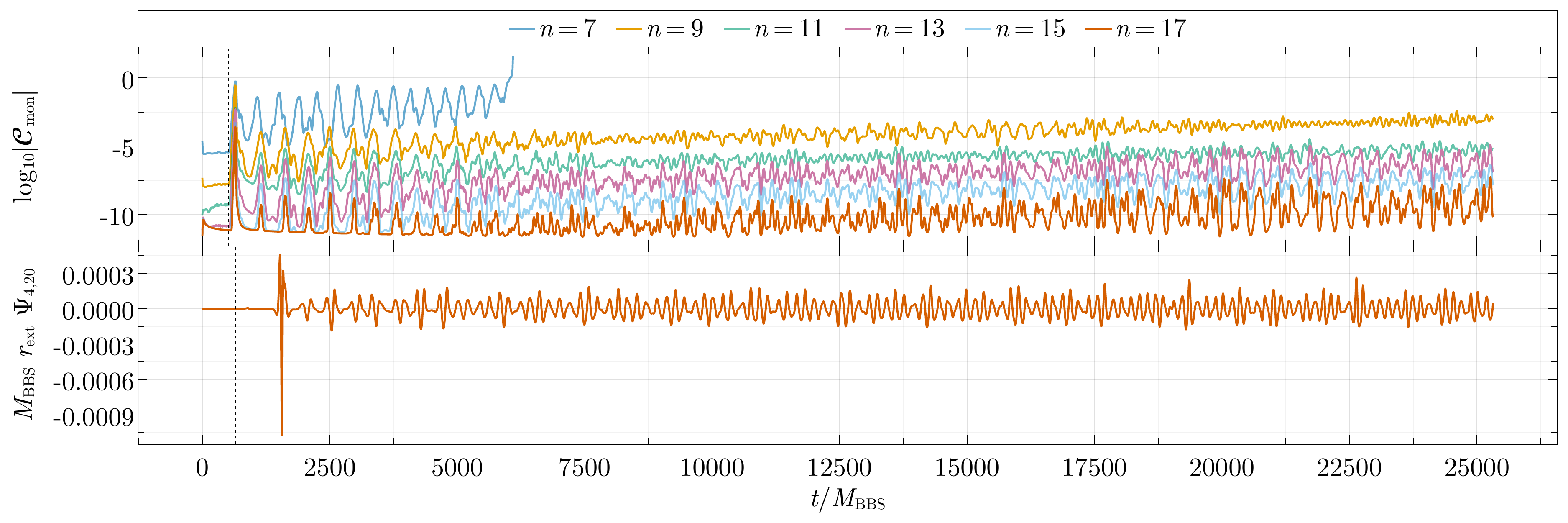}
  \caption{
    Time evolution of $\Cmon$ (top) for simulations based on CTS+CVE data
    with boost $v=0.05$ and initial separation $d = 80$
    for different resolutions $n$.
    Unprocessed $\PsiFT$ signal (bottom) obtained from the simulation with $n=17$
    and extracted on a sphere of radius $\Rext = 720$.
    The vertical dashed line indicates the time of merger
    as determined by the time of the maximum value of $\Amax(t)$ from the highest resolution.
  }
  \label{fig:cmon_psi4_longrun}
\end{figure*}

\autoref{fig:convergence_totalM_noetherQ} displays the time evolutions
of $N$, $\Mtot$ and $\MADM$ (top)
as well as a self-convergence study of the same
quantities (bottom).
These results were obtained from evolutions
of runs done with the CTS+CVE data with initial separation $d = 80$ and boost parameter $v = 0.05$.
Focusing first on the upper panel and the top plot which shows $N(t)/N(0)$
one can see that for resolution $n = 13$
the relative deviation from the initial value $N(0)$ is well below $1\%$
when $ 0 \leq t \leq 2000$, indicating that $N(t)$ is numerically conserved.
On the other hand, for $t \gtrsim 2000$ the Noether charge starts to decrease
until $\approx 98\%$ of the initial value is left at $t = 4000$.
Similar behavior is observed in the evolution of $\MADM(t)/\MADM(0)$,
\ie{} the ADM mass is conserved well below $1\%$ over the time range $ 0 \leq t \leq 2000$,
after which it starts to decrease until it reaches a final value below $97\%$ at $t = 4000$.
On the other hand, $\Mtot(t)/\Mtot(0)$ is by definition not a conserved quantity and this is
reflected in its time evolution, because it shows significant oscillations
already for $t \leq 2000$.
Note also that this quantity shows a decreasing trend for late times.
We verified that the apparent \textit{loss} in $N$, $\MADM$ and $\Mtot$ is due to the scalar field
getting absorbed by the outer boundary conditions,
much like the artificial oscillations of $A$ that we discussed in~\autoref{fig:amppsi_2d}.

Regarding the self-convergence tests displayed in the lower panel,
the differences in $\Delta N$ decrease with increasing resolution,
confirming that $N$ is indeed convergent.
We note that the differences are noticeably increased through the merger, but overall
remain roughly constant in time.
The differences in $\Delta \MADM$ show more variations in time, but
one can still recognize a convergence trend in this data.
We attribute this behavioral difference to the fact that $N$ and $\MADM$ are computed through
a volume and surface integral, respectively, and the latter being more susceptible to
numerical errors, because its computation involves less degrees of freedom
than that of a volume integral.
The differences in $\Delta \Mtot$ also behave similarly to $\Delta N$
in the way that they show a convergence pattern, the differences
remain roughly constant over time and they are increased through the merger.
These plots demonstrate that also global quantities like $\Delta N$ and $\Delta \Mtot$
converge exponentially.

\subsection{Gravitational waves}

\autoref{fig:cmon_psi4_longrun} shows results of a long-time evolution of
a BS head-on with initial separation $d = 80$ and initial boost parameter $v = 0.05$.
The plot shows that the run with the lowest resolution $n=7$ crashed,
but starting with $n\geq 9$ a stable convergence pattern can be observed where the exponential
improvement between resolutions is maintained also for late times.
One can also recognize a growth of $\Cmon$ over time, which might become troublesome
for even longer evolutions; however, we were able to evolve to at least $t/\MBBS = 50000$ without
problems and there remains the possibility to adjust the GHG damping system through the
parameter $\gamma_0$ in combination with adjustments to the local grid widths.
The $\PsiFT$ strain extracted from the highest resolved run with $n=17$ displays
the characteristic afterglow signature of these kinds of collisions, where
the remnant scalar-field cloud is continuing to oscillate and emit gravitational radiation
with an amplitude that is of the same order as the merger spike~\cite{croft2023gravitational}.
The dominant amplitude in the frequency spectrum of $\PsiFT$ is located at
\begin{align*}
  f_{\textrm{dom}} \approx 7.7 \times 10^{-3} \left( \frac{\mu}{6.582\times 10^{-16}~\text{eV}} \right)~\text{Hz} \, .
\end{align*}
Assuming a scalar-field mass $\mu = 1 \times 10^{-11}~\text{eV}$
this translates into $f_{\textrm{dom}} \approx 117~\text{Hz}$.
Furthemore, the plot shows that the afterglow lasts at least for time
\begin{align*}
  T_{\textrm{glow}} \gtrsim 18750 \left( \frac{\mu}{6.582\times 10^{-16}~\text{eV}} \right)^{-1}~\text{s} \, ,
\end{align*}
which corresponds to $T_{\textrm{glow}} \gtrsim 1.2~\text{s}$, assuming the same scalar-field mass.

We want to highlight that our computational setup does not involve angular momentum,
because the head-on collisions proceed with zero impact parameter and the simulations
are carried out in axisymmetry and with reflection symmetry, and yet we do observe
a noticable afterglow signature in \autoref{fig:cmon_psi4_longrun}.
Our results therefore demonstrate that the emission of GW afterglow radiation
in BS collisions is not solely tied to the presence of angular momentum in the initial data
or the remnant.

\begin{figure*}[t]
  \centering
  \includegraphics[width=\textwidth]{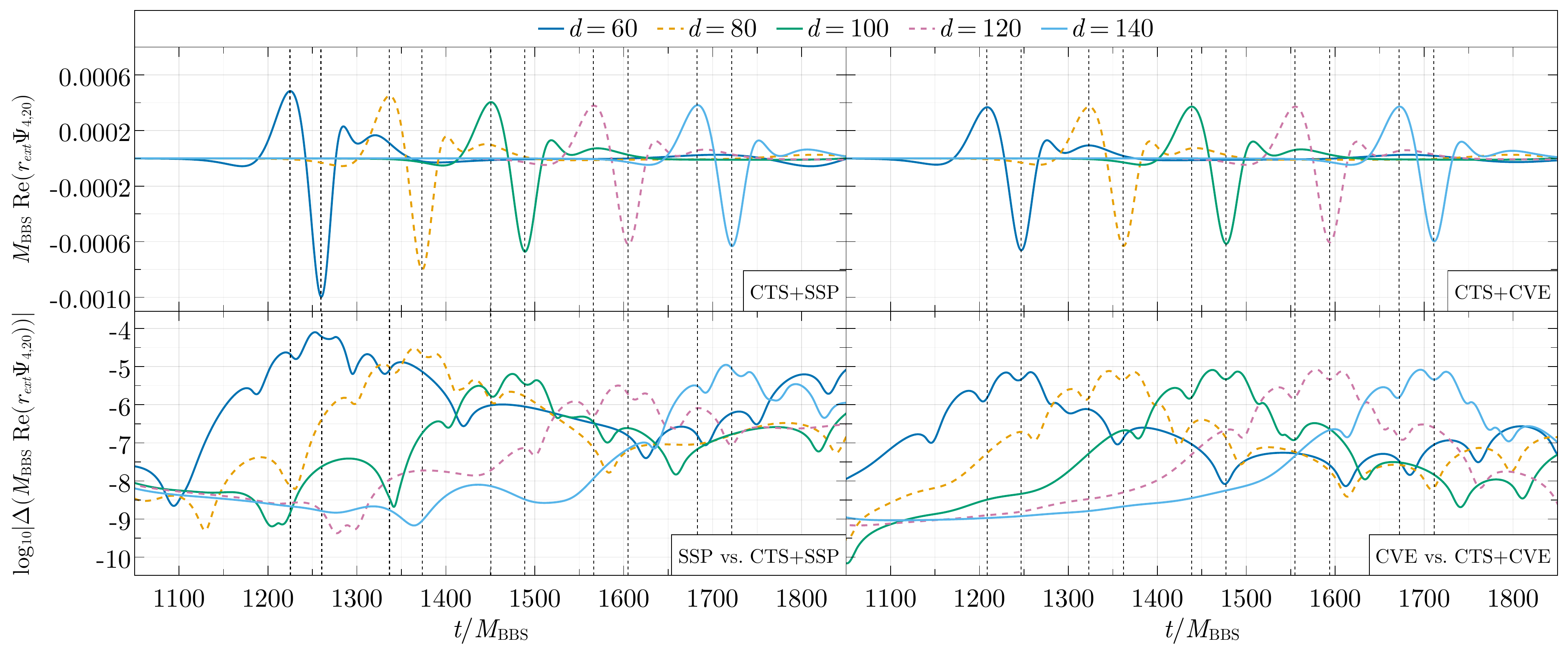}
  \caption{
    \textit{Top} $\PsiFT$ signals obtained from evolutions of CTS+SSP initial data (left)
    and CTS+CVE initial (right) with polynomial resolution $n=15$, initial boost parameter $v=0.1$
    and for varying initial distance $d$.
    \textit{Bottom} Differences of $\PsiFT$ signals between SSP \vs{} CTS+SSP (left)
    data and between CVE \vs{} CTS+CVE data (right).
    The $\PsiFT$ data was extracted on a sphere of radius $r_{\textrm{ext}} = 720$.
    The vertical dashed lines indicate the times where the extrema of $\PsiFT$ appear
    for each separation $d$.
  }
  \label{fig:psi4_compare_v_d}
\end{figure*}

In~\cite{helfer2022malaise} a comparison between $\PsiFT$ signals extracted around the time
of merger and obtained from runs using SSP and CVE initial data was already carried out.
The binary BS head-on configuration studied in there used the same isolated BS solution as
we do for the superposition. The initial boost parameter was fixed to $v = 0.1$.
In \autoref{fig:psi4_compare_v_d} we reproduce their results for resolution $n=15$, but using instead
CTS+SSP initial data (top, left) and CTS+CVE initial data (top, right).
Furthermore, we also evolved SSP and CVE initial data and we plot
the absolute differences between the $\PsiFT$ signals
obtained from SSP \vs{} CTS+SSP data (bottom, left) and
CVE \vs{} CTS+CVE data (bottom, right).
The data was extracted at spheres with radius $\Rext = 720$.
Focusing on raw $\PsiFT$ data first (top, left and right), one can observe
the same qualitative differences between CTS+SSP and CTS+CVE data
that were reported for SSP and CVE data in~\cite{helfer2022malaise}.
The maximum amplitude of $\PsiFT$ shows a dependence on the initial separation $d$ between the two stars
for the runs that used CTS+SSP data, whereas the CTS+CVE results show almost constant
maximum amplitude and the wave's shape is roughly independent of $d$.
Furthermore, for $d=80$ the GW signal arrives later by $\Delta t/\MBBS \approx 14$ from
CTS+SSP data compared to the CTS+CVE data and this time delay decreases
to $\Delta t/\MBBS \approx 10$ for $d=140$.
From the plot showing the differences of $\PsiFT$,
one can read off that the GW signal differs at most by $10^{-4}$
for SSP \vs{} CTS+SSP data and at most by $10^{-5}$ for CVE \vs{} CTS+CVE data.
When normalized by the maximum amplitude of the $\PsiFT$ signal,
these differences translate into a maximum deviation of $\approx 10\%$
for SSP \vs{} CTS+SSP and
$\approx 2\%$ CVE \vs{} CTS+CVE data, respectively.

\begin{figure}[t]
  \centering
  \includegraphics[width=0.99\columnwidth]{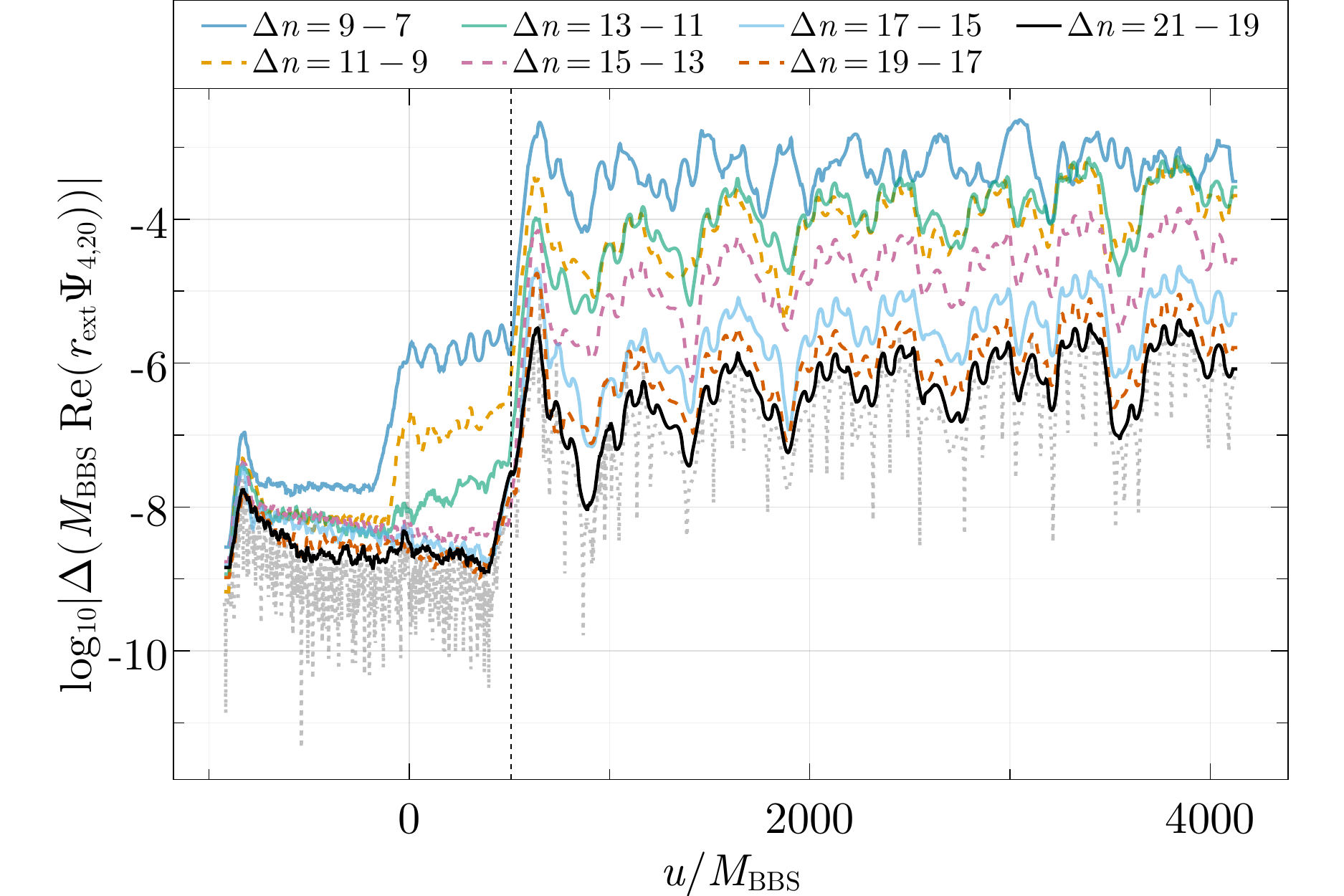}
  \caption{
    Self-convergence test of $\MBBS \Rext \PsiFT$
    obtained from evolutions of CVE data with initial separation $d = 80$ and boost parameter $v = 0.05$.
    Signals were extracted at coordinate radius $\Rext = 720$.
    The solid and dashed lines display differences between two consecutive resolutions of $\Psi_{4,20}$.
    For visualization purposes the data has been interpolated onto a common time grid for
    the difference computation and then postprocessed using a moving
    average filter (see \autoref{subsection:moving-average}).
    The dotted gray line is the raw difference computed between resolutions $n= 19$ and 21,
    which illustrates that the moving average accurately represents the trend of the data.
    The vertical dashed line indicates the time of merger
    as determined by the time of the maximum value of $\Amax(t)$ from the highest resolution.
  }
  \label{fig:psi4_self-convergence}
\end{figure}

\autoref{fig:psi4_self-convergence} presents a self-convergence test using $\PsiFT$
signals extracted from evolutions of CTS+CVE initial data
with initial separation $d = 80$ and boost parameter $v = 0.05$.
The signals were extracted at spheres with coordinate radius $\Rext = 720$
and the results are plotted against retarded time $u$.
Similar to the self-convergence test of the global quantities $N$, $\Mtot$ and $\MADM$,
one can observe a clear decrease in the differences $\Delta \Psi_4$ with
increasing polynomial resolution, indicating an exponentially convergent result.
The differences increase notably through the merger by roughly two orders
of magnitude.
The differences between the two highest resolved runs with $n=17$ and $19$ can
be interpreted as an error bound on the numerical accuracy of $\PsiFT$
and we infer that the results are accurate up to $10^{-5}$ for the studied configuration.
This bound is conservative, because, as the trend of this convergence test indicates,
any differences obtained between two consecutive resolutions that are each higher than $n=19$
will be smaller than this difference.

\begin{figure}[t]
  \centering
  \includegraphics[width=0.99\columnwidth]{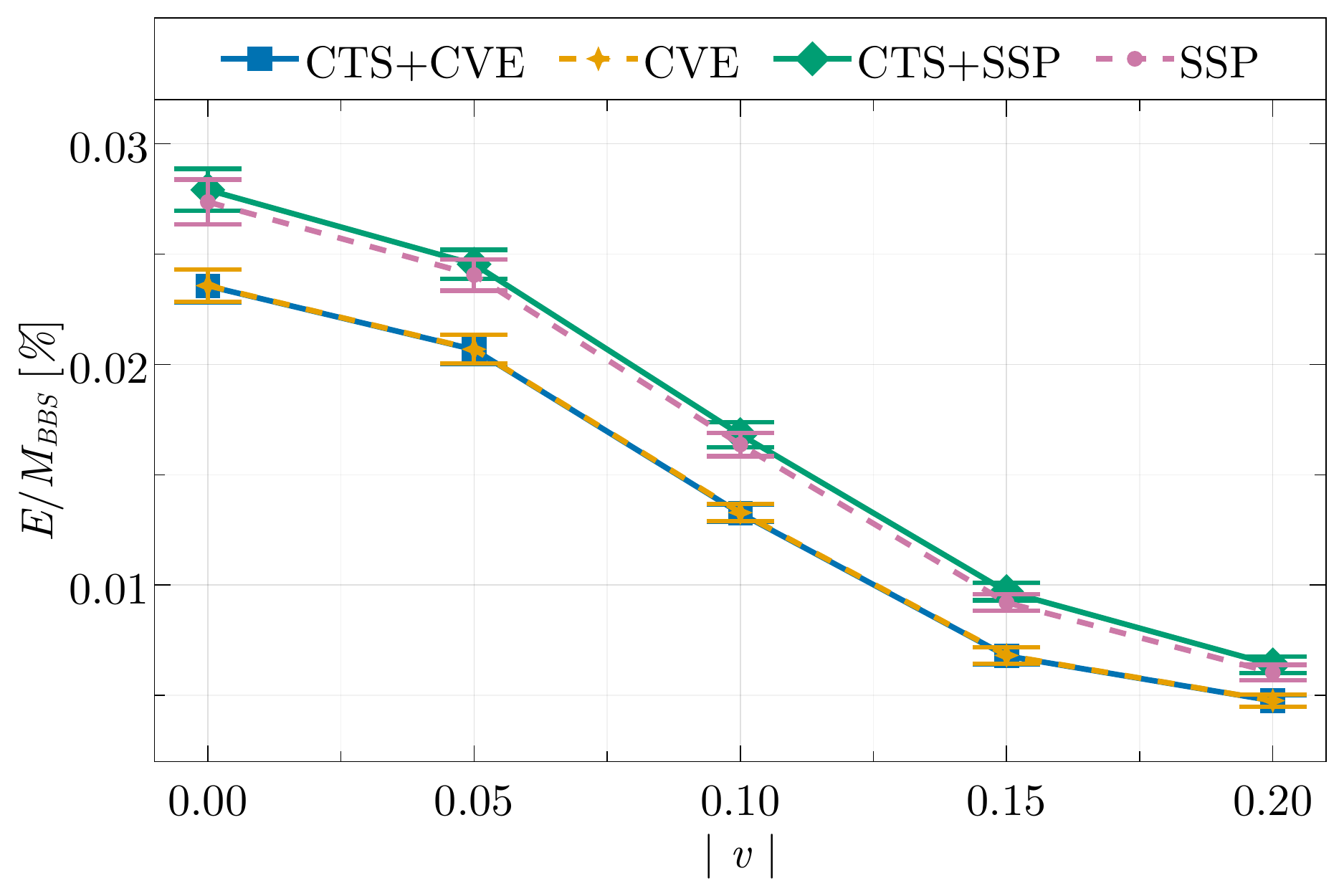}
  \caption{
    Dependency of the radiated GW energy $E/\MBBS$ on the initial boost parameter $v$
    obtained from evolutions of SSP, CTS+SSP, CVE and CTS+CVE initial data
    and fixed initial distance $d = 80$ and polynomial resolution $n=15$.
    The error bars account for finite resolution errors and
    uncertainties in the postprocessing.
  }
  \label{fig:energy-vs-v}
\end{figure}

\begin{figure}[t]
  \centering
  \includegraphics[width=0.99\columnwidth]{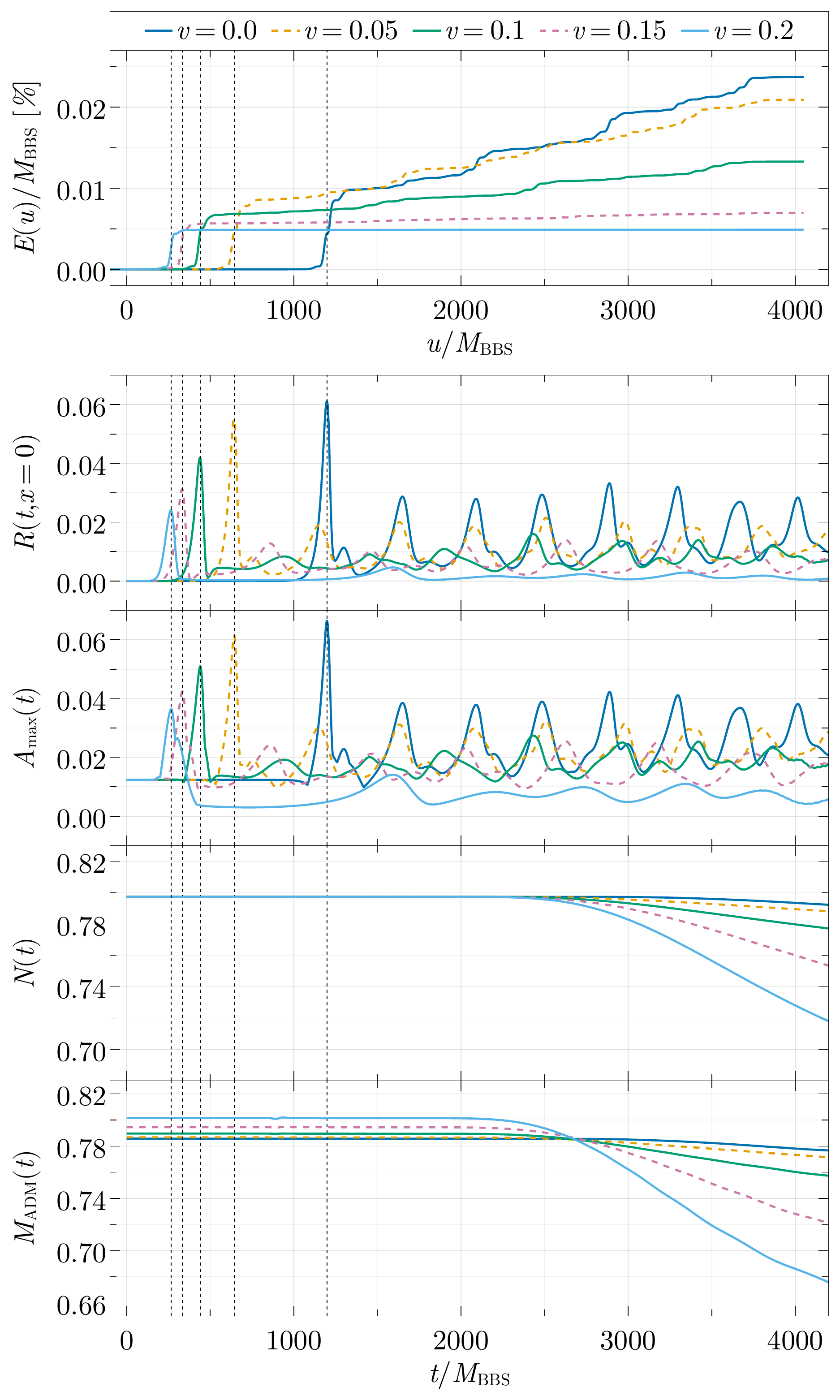}
  \caption{
    \textit{Top} Time evolution of $E(u)/\MBBS$ \vs{} retarded time $u/\MBBS$
    obtained from evolutions of CTS+CVE initial data with fixed initial separation
    $d= 80$ and resolution $n=15$, but varying boost parameter $v$.
    \textit{Bottom} Time evolution of $R(t,x=0)$, $\Amax(t)$, $N(t)$,
    and $\MADM(t)$ \vs{} coordinate time $t/\MBBS$ from the same simulations.
    Note that $E/\MBBS$ stops growing when $u/\MBBS \gtrsim 3800$ for all values of $v$,
    which is due to the use of a window function in the postprocessing of the data
    (see \aref{Appendix:gw-analysis} for details).
    The vertical dashed lines indicate the time of merger
    as determined by the time of the maximum value of $\Amax(t)$.
  }
  \label{fig:energy-vs-time}
\end{figure}

In \autoref{fig:energy-vs-v} we compare the radiated gravitational wave energy $E/\MBBS$
emitted over a retarded time span of $u \in [100, 3000]$
from evolutions of SSP, CTS+SSP, CVE and CTS+CVE initial data,
where we fix the initial separation $d = 80$ and polynomial resolution $n=15$ and vary the
boost parameter $v$.
First, the plot shows a difference in $E$
depending on whether SSP or CVE initial data is evolved and this difference decreases with
increasing $v$.
On the other hand, the difference between $E$ obtained from SSP \vs{} CTS+SSP
and CVE \vs{} CTS+CVE is comparatively negligible.
Taking into account the error bars the conclusion is that one cannot distinguish
superposed initial data from constraint-solved data by the amount of radiated energy
emitted over this time span.
Regarding the dependence on $v$, it is evident that $E$ decreases with
increasing $v$. From a physical perspective this is counter intuitive,
because increasing $v$ increases the energy of the binary system in the initial slice.
Furthermore, with larger $v$ the time to merger is reduced and, thus, there would
be more time left for GWs to radiate away from merger till $u=3000$.
This behavior is in contrast with a similar experiment that was conducted for
BH head-on collisions in \cite{sperhake2008high}, where indeed the
radiated GW energy increases when increasing $v$.
Nevertheless, these results are in accordance with what was observed in~\cite{helfer2022malaise},
where $E$ also decreases when the initial separation $d$ between the stars is increased,
which also corresponds to an increase of the total energy in the initial slice.

Analysis of 1D and 2D output from simulations that were used for \autoref{fig:energy-vs-v}
shows that with increasing $v$ the gravitational and scalar field both display decreasing amplitudes.
To this end, consider \autoref{fig:energy-vs-time} where the time evolution of $E(u)/\MBBS$,
$R(t,x=0)$, $\Amax(t)$ $N(t)$ and $\MADM(t)$ is shown for simulations of CTS+CVE data with
initial separation $d= 80$ and resolution $n=15$, but we vary the boost parameter $v$.
First, one can read off the values of $E/\MBBS$ at $u/\MBBS \approx 4050$ which were shown
in \autoref{fig:energy-vs-v} for the CTS+CVE data.
The plot displays that the initial burst of radiation is responsible for a significant increase in $E/\MBBS(u)$.
When contrasted with the evolutions of $R(t,x=0)$ and $\Amax(t)$, it is apparent that this is
caused by the merger where these quantities peak.
Furthermore, one observes that with increasing $v$ the first burst is emitted at earlier times,
but the strength of the burst also decreases.
However, at the same time the flux of $E(u)/\MBBS$ radiated in the afterglow phase decreases when $v$ increases.
During that phase $R(t,x=0)$ and $\Amax(t)$ also display a continued decrease in the amplitude
of the oscillations.
The evolutions of $N(t)$ and $\MADM(t)$ show that until $t/\MBBS \approx 2000$
the quantities $N$ and $\MADM$ are both conserved to good approximation.
The difference in $\MADM$ when $0 \leq t/\MBBS \leq 2000$ between different values of $v$
reflects the fact that an increase in the initial momenta of the stars
corresponds to initially more energetic configurations.
Eventually, both $N(t)$ and $\MADM(t)$ decrease, because the scalar field
leaves the computational domain through the boundary.
We note that this decrease sets in earlier and proceeds faster when $v=0.2$
than $v=0.0$, which we verified by studying $dN/dt$ (not shown).
This allows one to conclude that the larger $v$, the faster energy is carried
away through the scalar field being ejected. As a consequence, the produced remnant will be lighter for larger $v$.

\begin{figure}[t]
  \centering
  \includegraphics[width=0.99\columnwidth]{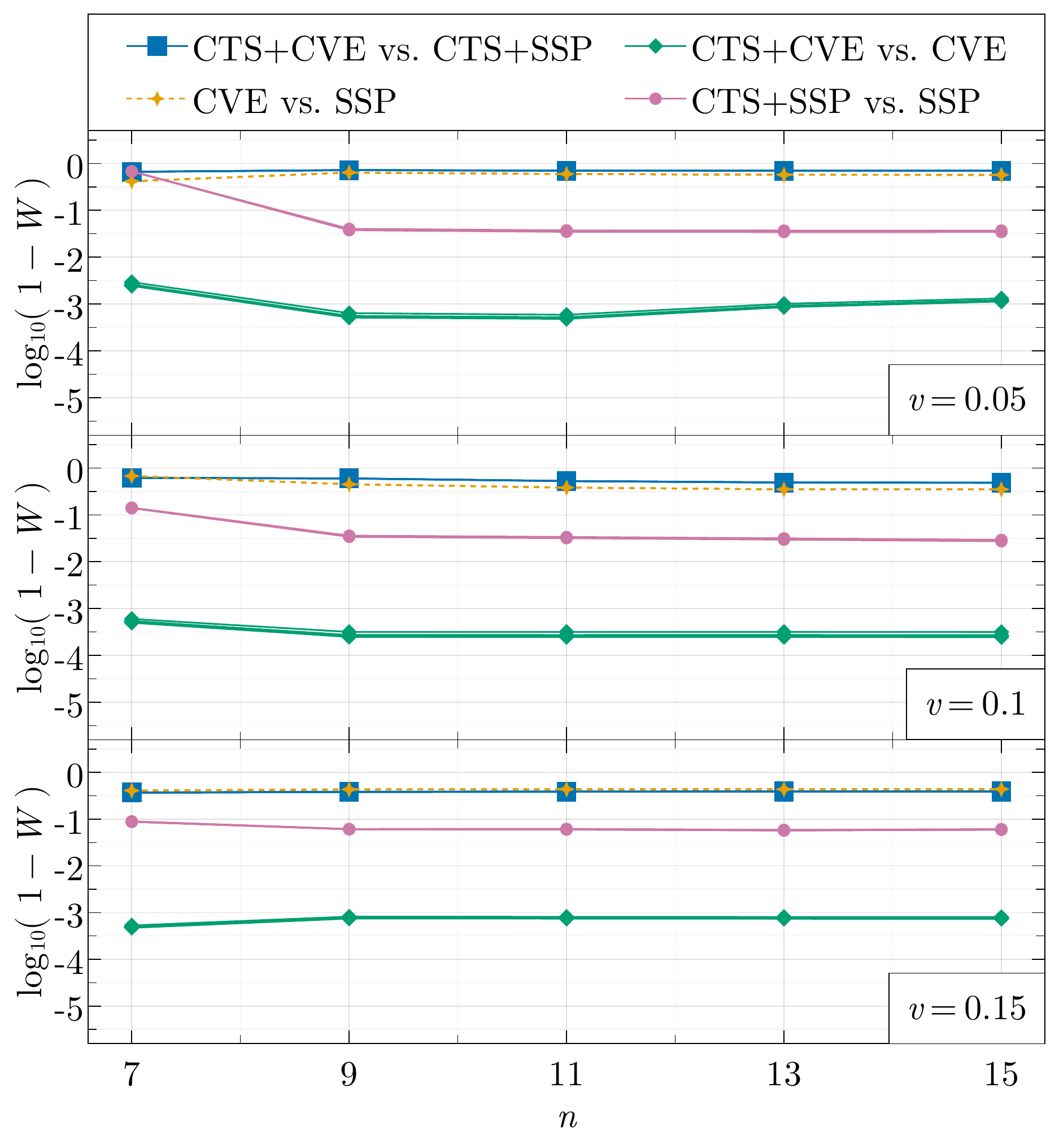}
  \caption{
    Analysis of the complementary Wiener product $\bar W(\PsiFT^{(1)},\PsiFT^{(2)})$
    between GWs $\PsiFT^{(1)}$ and $\PsiFT^{(2)}$ obtained from two initial data sets
    and for varying numerical resolution $n$.
    The initial stars were prepared with a separation $d = 80$ and boost parameters
    $v = 0.05$ (top), $v = 0.1$ (middle), $v = 0.15$ (bottom).
  }
  \label{fig:wiener-product-vs-n}
\end{figure}

\autoref{fig:wiener-product-vs-n} shows the (complementary) Wiener product
$\bar{W} := 1 - W({\PsiFT}^{(1)}, {\PsiFT}^{(2)})$
computed between two signals ${\PsiFT}^{(1)}$ and ${\PsiFT}^{(2)}$ and
obtained from different initial data constructions.
The initial distance was fixed to $d = 80$, we vary the polynomial resolution $n$
and we show results for boost values $v= $ 0.05, 0.1 and 0.15.
We used the \texttt{T1800044-v5} Advanced LIGO sensitivity design curve~\cite{ligopsd}
to weight the product and the complex scalar field's mass was fixed to $\mu = 1\times10^{-11}~\text{eV}$.
$W$ was computed using data that was extrapolated to null infinity with different orders $\Nextrp$.
We remind the reader that a value $\bar{W} = 1 - W({\PsiFT}^{(1)}, {\PsiFT}^{(2)}) = 0$ means
that ${\PsiFT}^{(1)}$ and ${\PsiFT}^{(2)}$ are identical if they would appear in
a detector with the provided noise sensitivity curve.
The plot shows, independent of $v$, that $\bar{W}$ is always smallest
when comparing $\PsiFT$ from evolutions of CVE data \vs{} CTS+CVE data,
meaning that these waveforms are the closest to each other among those we compare.
The difference between GWs received from SSP data \vs{} CTS+SSP data is bigger than
the one from CVE data \vs{} CTS+CVE data.
$\bar{W}$ is biggest when comparing either SSP data \vs{} CVE data or CTS+SSP data \vs{} CTS+CVE data.
We note that these qualitative results are to very good agreement independent
of the numerical resolution $n$.
Furthermore, the results are independent of the extrapolation order $\Nextrp$,
because \autoref{fig:wiener-product-vs-n} already shows the results obtained from analyses where
$\Nextrp$ was varied too, however, the obtained curves all overlap with one another.
Overall this picture is consistent with the discussions of other comparisons
in this work where it turned out that any apparent differences
between analysis quantities are more pronounced when comparing results from
SSP data \vs{} CTS+SSP data than comparing results from CVE data \vs{} CTS+CVE data.

\section{Conclusion}
\label{Section:Conclusion}

In this manuscript we reported results of long-time stable and accurate
binary head-on BS collisions obtained using a PS method as implemented in \bamps{}.
We studied variations of a mini BS configuration which was already investigated
in~\cite{helfer2022malaise}, but differed in the way the initial data was constructed:
two of the data sets (SSP, CVE) were obtained from a superposition of isolated stars and thus
carried initial constraint violations, whereas the other two data sets (CTS+SSP, CTS+CVE) were obtained
by numerically solving the CTS equations and with free data based on SSP and CVE data,
using the hyperbolic relaxation method presented in~\cite{ruter2018hyperbolic}.
In our work we did not enforce quasi-equilibrium conditions to ease our comparison
with previous work in~\cite{helfer2022malaise}.
This effort was undertaken to answer the question of how much of a difference one can
expect physical observables to vary when numerically evolving constraint-violating and
satisfying data.

In summary we found that the differences in the discussed analysis quantities computed
during evolutions, among which is also the constraint monitor of \bamps{},
are always bigger between SSP \vs{} CTS+SSP data
when compared to CVE \vs{} CTS+CVE data.
We demonstrated that evolutions of SSP and CVE data bear residual constraint violations
above $10^{-8}$ and $10^{-10}$, respectively,
despite being self-convergent.
On the other hand, we could reduce and preserve the constraint violations for CTS+SSP and
CTS+CVE data well below $10^{-10}$ using a reasonable resolution,
which demonstrates the capabilities of PS approximations as viable methods
for next generation NR codes targeted at smooth solutions.
A study of global analysis quantities showed that
conserved variables like Noether charge $N$ and ADM mass $\MADM$ can be preserved well below $1\%$ of their
initial value already starting with comparatively small polynomial order $n=9$ per grid,
as long as the scalar field does not leave the computational domain, while
at the same time also displaying exponential decrease in a self-convergence study.

A direct comparison of the Ricci scalar at the collision center $R(x=0)$ and the
scalar field amplitude $\Amax$ between results obtained from
SSP \vs{} CTS+SSP showed that these quantities eventually start to dephase after merger.
No such deviations were observed when comparing results
from evolutions of CVE \vs{} CTS+CVE initial data.
The main motivation for the development of the CVE method in~\cite{helfer2022malaise} was
that for head-on collisions of solitonic BSs $\Amax$ obtained from SSP data displayed
an artificial growth during the infall phase and eventually lead to premature BH collapse.
This growth was avoided by the use of the CVE construction, which also delayed the BH formation
after the stars' centers merged.
Although, this work was concerned with mini BS collisions for which no BH collapse
occurs, we find a qualitatively similar behavior for $\Amax$ when evolving SSP and CVE data.
Furthermore, the constraint-solved CTS+SSP data did not cure this artificial growth pre-merger.
On the other hand, the evolution of $\Amax$ obtained from CTS+CVE data
is to good approximation identical to the CVE data, \ie{} $\Amax$ remains approximately
constant during infall when the centers of the stars are well separated.
This latter observation serves as an argument which favors the use of CVE and CTS+CVE data
over SSP and CTS+SSP data, in particular, when the goal is to model the coalescence
of initially non-interacting stars that fall in on each other.

We reproduced the $\PsiFT$ signals of the mini BS head-on collisions that were reported in~\cite{helfer2022malaise}
for SSP and CVE data and compared these findings with signals obtained from CTS+SSP and CTS+CVE data.
We found that constraint solving does not significantly alter the
qualitative differences that were discussed in~\cite{helfer2022malaise}.
These results are robust, as we demonstrated exponential decrease in
differences of $\PsiFT$ signals in a self-convergence study.
We then computed the radiated energy from these kinds of head-on collisions while also varying
the initial boost parameter $v$ and found, again, a difference that is dominated by
the way in which the isolated stars are superposed.
For this quantity differences due to constraint solving the initial data are negligible
and indistinguishable when accounting for errors in postprocessing.
The analysis of the errors required special attention
due to the lack of a heuristic cutoff frequency for signals originating from head-on collisions
and which is used to suppress nonlinear drifts in the GW reconstruction
(see \aref{Appendix:gw-analysis}).

An interesting physical result is that the radiated energy emitted during a fixed interval of
coordinate time decreases when the initial momenta of the stars is increased,
which is in opposition to BH head-on collisions where increasing the
progenitors initial momenta leads to more energy being radiated gravitationally~\cite{sperhake2008high}.
However, these findings are conceptually in agreement with results presented in~\cite{helfer2022malaise}
where for fixed boost parameter $v$ the radiated energy
also decreases when the initial distance between two BSs is increased, which also corresponds
to an increase of the total energy of the system.
This qualitative behavior is independent of the quality of the initial data, but
quantitative differences in total radiated energy could be observed.

We quantified the difference in the shape of the GW signals by computing the
Wiener product between results of SSP, CVE, CTS+SSP and CTS+CVE initial data and varying resolutions,
while taking into account detector noise.
Here we found that signals obtained from CVE \vs{} CTS+CVE data are more similar to one another
than signals obtained from SSP \vs{} CTS+SSP data.
Comparing signals between SSP \vs{} CVE and CTS+SSP \vs{} CTS+CVE showed that the way in which the superposition
of stars is done dominates any effects due to constraint violations, which is
in agreement with the rest of our findings.

Overall we can conclude that all the differences we observed in the analysis quantities
we studied, and for the particular BS head-on configuration we considered, are
dominated by the way in which the SSP and CVE constructions differ.
Differences in physical observables due to constraint violations were comparatively negligible,
and marginally relevant when accounting for errors.
This result is reassuring for theoretical predictions that were made based
on BS simulations that used superposed data and for which it was verified
that the initial constraint violations were reasonably small.
Nevertheless, we recommend the use of constraint-satisfying over constraint-violating
initial data for multiple reasons.
First, solutions to the EFEs satisfy the constraint equations exactly at all times,
which includes the initial slice, and, thus, the goal for numerical approximations
of these solutions should be to satisfy these equations as well as possible.
In practice it is also easier to preserve small initial constraint
violations in time than having to rely on large initial constraint violations being reduced in time
through a damping scheme, or having to rely on them to leave the computational domain through a boundary.
Furthermore, it is not uncommon in long-time evolutions to find
that constraint violations eventually start to grow for late times.
In such a scenario, constraint-satisfying data will likely allow to evolve for
longer times than constraint-violating data would, because the former leaves more room
for the growth of violations during the evolution until the result becomes dominated by errors
or the simulation crashes.

This work established the basis for future studies of BS evolutions with the \bamps{} code.
With the addition of AMR support in \bamps{}~\cite{renkhoff2023adaptive} and the nice convergence
behavior demonstrated in this work, we are confident that our code will be able to
perform high-resolution and long-time-stable simulations of binary inspiraling BS collisions
to produce quality GW signals. A future goal is to eventually build a waveform template bank
for a targeted search of merger signals involving exotic compact objects.
Another avenue worth exploring would be the problem of the construction of quasi-equilibrium
initial data for which first work has started in~\cite{siemonsen2023generic}.
In particular the hyperbolic relaxation method used to solve the CTS equations in the present paper
is directly implemented in \bamps{} and, as such, it can profit from all optimizations
that are added to the evolution code, which should enable generation of high quality initial data
with a moderate amount of computational resources.

\begin{datastatement}

The data that support the findings of this study will be made available in the
\textsc{CoRe}~database~\cite{gonzalez2023second,core-website}.

\end{datastatement}

\begin{acknowledgments}

We thank Rossella Gamba, Thomas Helfer, Robin Croft and Ulrich Sperhake
for providing valuable feedback for the manuscript.
We are also grateful to Thomas Helfer, Robin Croft and Ulrich Sperhake for answering questions
regarding their simulations, their GW radiated energy computations and for providing
access to their data.
F.A. is also thankful to Alexander Jercher for discussions on natural units in GR
and to Rosella Gamba for discussions on GW analysis
as well as pointing us to reference~\cite{bustillo2022gravitational}, which
lead to the addition of the Wiener product analysis.

Computations were performed on the ARA cluster at the Friedrich-Schiller University Jena
and on the supercomputer SuperMUC-NG at the Leibniz-Rechenzentrum (LRZ) Munich
under project number \texttt{pn36je}.

F.A., D.C. and R.R.M. acknowledge support by the
Deutsche Forschungsgemeinschaft (DFG) under Grant
No. 406116891 within the Research Training Group RTG 2522/1.
R.R.M. acknowledges support by the DFG under Grant No. 2176/7-1.
H.R.R. acknowledges
support from the Funda\c c\~ao para a Ci\^encia e Tecnologia (FCT)
within the Projects No. UID/04564/2021, No. UIDB/04564/2020, No. UIDP/04564/2020
and No. EXPL/FIS-AST/0735/2021.
We acknowledge financial support provided under the European Union’s H2020 ERC Advanced Grant
``Black holes: gravitational engines of discovery'' Grant Agreement No. Gravitas–101052587.

The figures in this article were produced with \textsc{Makie.jl}~\cite{danisch2021makie,makie-github},
\textsc{ParaView}~\cite{ayachit2015paraview,ahrens200536}, \textsc{Inkscape}~\cite{Inkscape}.

\end{acknowledgments}
\appendix
\section{Gravitational wave analysis}
\label{Appendix:gw-analysis}

Below we outline the GW analysis used in \bamps{} and describe the
postprocessing techniques used for the analysis of the data presented in the main text.

The \bamps{} code employs the Newman-Penrose formalism~\cite{newman1962approach} to compute the
curvature pseudo-scalar field $\Psi_4$, which measures radially outgoing gravitational radiation.
The actual implementation follows the standard procedure outlined
in~\cite{alcubierre2008introduction} (see also~\cite{bishop2016extraction} for a review),
where $\Psi_4$ is first computed from the Weyl tensor and an orthonormal null tetrad,
which itself is constructed from an orthonormal basis adapted to the normal direction
of spheres centered around the collision region. %
Taking into account the pseudo-scalar
characteristic of $\Psi_4$, it is then decomposed into a modal expansion
of spin-weighted spherical harmonics ${}_{-2}Y_{lm}$
with complex coefficients $\Psi_{4,lm}$, where $l \geq 2, m = -l, \dots, l$.

For the analysis of head-on collisions, which take place in axisymmetry,
only the expansion coefficients with $l$ even and $m = 0$ are non-zero, and are purely real~\cite{sperhake2008high}.
In this work we only analyze the dominant $l,m = 2,0$ mode.

The GW flux and strain is reconstructed from~\cite{alcubierre2008introduction}
\begin{align}
  \frac{\dd r h_{20}}{\dd t}(t) &= \lim_{r\to\infty} \int^{t}_{-\infty} \dd t' ~ r \Psi_{4,20}(t',r) \, ,
  \label{eq:instant-freq}
  \\
  r h_{20}(t) &= \lim_{r\to\infty} \int^{t}_{-\infty} \dd t' \int_{-\infty}^{t'} \dd t''~ r \Psi_{4,20}(t'',r) \, ,
  \label{eq:strain}
\end{align}
and the radiated energy associated with $\PsiFT$ is computed from
\begin{align}
  E(t) &= \int^t_{-\infty} \dd t'~ \frac{\dd E(t')}{\dd t} \, ,
  \label{eq:energy}
  \\
  \frac{\dd E}{\dd t}(t) &= \lim_{r\to\infty} \frac{1}{16\pi} \left| \int^t_{-\infty} \dd t'~ r\Psi_{4,20}(t',r) \right|^2 \, .
  \label{eq:energy-flux}
\end{align}

\begin{figure*}[t]
  \centering
  \includegraphics[width=\textwidth]{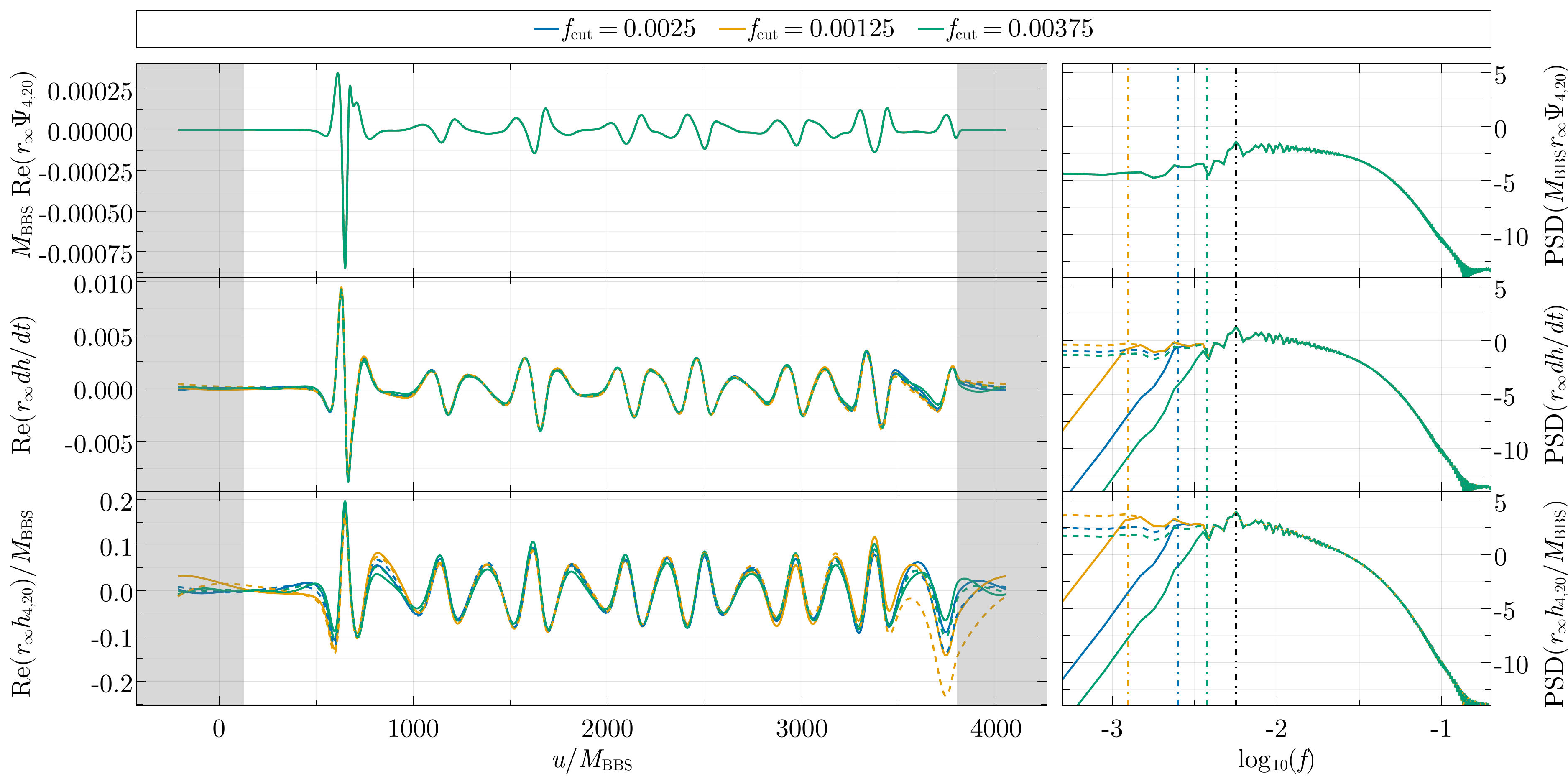}
  \caption{
    $r_\infty\PsiFT$ (top) signal and its PSD
    extrapolated to null infinity with $\Nextrp = 3$ and data obtained
    from a simulation of CTS+CVE initial data with polynomial resolution $n=15$,
    initial separation $d = 80$ and boost parameter $v = 0.05$.
    Gray shaded areas indicate where the windowing~\eqref{eq:psift_windowed} was applied.
    $dh/dt$ (middle) and $h$ (bottom) are results obtained with the methods
    discussed in the text:
    modified FFI method (solid, with extra Butterworth high-pass filtering),
    standard FFI method (dashed).
    Vertically dashed-dotted lines correspond to the different cutoff frequencies used
    and vertically dash-double-dotted lines correspond to the frequency of the maximum amplitude
    in $\text{PSD}(\MBBS \Rext \PsiFT)$.
  }
  \label{fig:psi4-dhdt-h}
\end{figure*}

In order to carry out the limit $r \to \infty$ above
we utilize the peeling property of $\PsiF$~\cite{newman1962approach} and perform an extrapolation in radius $r$
before integration.
To this end, we introduce the retarded time coordinate
\begin{align}
  u(t) &= u_\ast(t) - t_\ast \, ,
  \nonumber\\
  &= t - r - 2 \MBBS \log\left(\frac{r}{2\MBBS}-1\right) - t_\ast \, ,
\end{align}
where $t_\ast$ is a constant that is added in order to correct the retarded Schwarzschild
time coordinate $u_\ast$ to account for a potential time dependence of the lapse
and the radial coordinate in the far field zone, as well as for the fact that the
$r$ used in our simulations is in general not an areal coordinate in our
gauge~\cite{taylor2013comparing}.
$t_\ast$ is determined numerically by aligning multiple $\Psi_{4,20}$ data streams
extracted at different radii $\Rext$ at the first zero before the merger spike.
We note that in~\cite{taylor2013comparing,boyle2009extrapolating} more sophisticated expressions
for determining $t_\ast$ were provided which can also account for departures of the
standard tortoise Schwarzschild coordinate.
Their method requires extra simulation output, but we found that numerically determining
$t_\ast$ in a postprocessing step can also improve the quality of the extrapolation
to null infinity to a satisfying degree, as we discuss next.

The so obtained time aligned $\PsiFT$ data streams recorded at $N$ different
extraction radii ${\Rext}_{,1} \leq \dots \leq {\Rext}_{,N}$ are then fitted to the ansatz
\begin{align}
  r\Psi_{4,20}(u) = r_\infty\Psi_{4,20}(u) + \sum_{n=1}^{\Nextrp} \frac{\alpha_n(u)}{r^n} \, ,
\end{align}
using the linear least squares method to determine the time dependent coefficients $r_\infty\Psi_{4,20}$ and $\alpha_n$.
The desired extrapolated curvature scalar for the evaluation
of~\eqref{eq:instant-freq}, \eqref{eq:strain} and~\eqref{eq:energy-flux}
is then given by $r_\infty\Psi_{4,20}$.
We note that the quality of the extrapolated result depends 1) on the
number and range of extraction radii ${\Rext}_{,N} - {\Rext}_{,1}$ on which data is recorded (the more, the better),
2) the extrapolation order $\Nextrp$,
and 3) the time alignment through the correction $t_\ast$.
All results involving $\PsiFT$ and presented in this work were obtained from simulations where
the outermost boundary was located at $R = 800$ and we recorded
$\PsiFT$ data on 48 spheres with equally spaced increments between
${\Rext}_{,1} = 200$ and ${\Rext}_{,N} = 800$.
The extrapolations to null infinity of $r_\infty\PsiFT$ were only used for
\autoref{fig:energy-vs-v}, \autoref{fig:energy-vs-time} and \autoref{fig:wiener-product-vs-n}, for
which we verified that the results are stable between extrapolations that used different orders
$\Nextrp = 1, \dots, 5$.

The last ingredient for the analysis is the numerical evaluation of time integrals appearing
in the reconstruction formulae.
To this end we first preprocess the $r_\infty\PsiFT$ data by
limiting it to a range $u \in [u_L, u_R]$ with
\begin{align}
  w(u,\bar{u},\sigma) &= \frac{1}{2} (\tanh( \sigma (u - \bar{u}) ) - 1) \, ,
  \\
  r_\infty\PsiFT'(u) &= r_\infty\PsiFT(u) \nonumber\\
    &\times w(u,u_L,\sigma) w(u,u_R,-\sigma) \, .
  \label{eq:psift_windowed}
\end{align}
We used $\sigma=1/10$, $u_L=100$ and $u_R=3000$ for all our analysis
to make the signal periodic in $u$,
which avoids additional artifacts when Fourier analyzing the signal later on.
We note that the choice of $u_L$ is such that the dominant spike in the analysis of all
studied GW signals is not affected by the windowing, even when the boost parameter is $v = 0.2$.
Given the windowed data, we then evaluate the integrals
using a variation of the fixed frequency method (FFI)~\cite{reisswig2011notes},
which requires a choice for the cutoff frequency $\fcut$ to remove spurious nonlinear drifts
and we comment on this further below.
In our variation to the FFI method we first apply a discrete Fourier transform (DFT) to the windowed
data to obtain $\mathcal{F}(r_\infty\PsiFT)$,
we then apply a high-pass Butterworth filter of order $N=10$ with transfer function $H_N(f,\fcut)$
to obtain
\begin{align}
  \mathcal{F'}(r_\infty\PsiFT) = \mathcal{F}(r_\infty\PsiFT) | H_N(f,\fcut) | \, ,
\end{align}
and finally perform the FFI integration with $\mathcal{F'}(r_\infty\PsiFT)$
and cutoff $\fcut$ to compute~\eqref{eq:instant-freq}, \eqref{eq:strain} and~\eqref{eq:energy-flux}.
Lastly, to compute $E$ via~\eqref{eq:energy} we resort to using a standard trapezoidal rule,
since $E(t)$ will be monotonically increasing and the FFI method can only be used
for oscillating signals.

\autoref{fig:psi4-dhdt-h} shows an example of an extrapolated and windowed $\PsiFT$ data stream,
reconstructions $\dd h_{4,20}/\dd t$ and $h_{4,20}$, as well as the respective
power spectral densities (PSDs), that were computed with three
different values for $\fcut$ (see legend).
The PSD of the $\dd h_{4,20}/\dd t$ results show that
the integration amplified small frequency components and
the additional processing via a Butterworth
high-pass filter exponentially suppresses this growth for frequencies below $\fcut$.
The choice of $\fcut$ appears to have a mild impact on the time domain signal
$\dd h_{4,20}/\dd t$.
On the other hand, the PSD of $h_{4,20}$ demonstrates that for too small values of $\fcut$
low frequency components are amplified to a level that they become comparable to
components above $\fcut$. This translates into a strong dependence
of the time domain signal on the choice of $\fcut$, even when using the Butterworth
filter.
From this test we conclude that $\dd h_{4,20}/\dd t$ is robust enough for the computations
of $\dd E/\dd t$ and $E$ via~\eqref{eq:energy-flux} and~\eqref{eq:energy},
whereas the quality of $h_{4,20}$ in our analysis is not sufficient for further studies
and we leave it to future work to improve on this.

In~\cite{reisswig2011notes} it was established that a good choice for
the cutoff frequency for the $(l,m)=(2,2)$ mode for binary inspiral simulations is given by
$\fcut \approx 2 f_{\textrm{orb}}$,
where $f_{\textrm{orb}}$ is the initial orbital frequency of the stars.
Unfortunately, we are not aware of a similar heuristic for head-ons due
to the absence of orbital motion and the time domain signal displaying a
dominant pulse due to the merger.
The three values of $\fcut$ given in \autoref{fig:psi4-dhdt-h} are
used for all computations of $E$
for which results are shown in the main text (including varying the boost parameter).
The associated error bars account for 1) finite resolution errors
and 2) uncertainties due to variation in the extrapolation order $\Nextrp$
and variation in $\fcut$.
To estimate those contributions we proceed as follows.
We compute $E$ for two resolutions $n=13$ and $n=15$ and
for all combinations of $\Nextrp$ and $\fcut$.
The contribution 1) is taken to be the maximum of all differences of $E$ between results of resolution
$n=13$ and $n=15$, but the same values of $\Nextrp$ and $\fcut$.
For contribution 2) we compute the average of all $E$ results obtained for $n=15$
and all combinations of $\Nextrp$ and $\fcut$. The associated error is then
defined as the maximum of all differences of $E$ and the average of $E$.
These two errors are then combined in quadrature and added in \autoref{fig:energy-vs-v}.
We note that the finite resolution error is on average two orders of magnitude
smaller than the error due to the uncertainties involved in the postprocessing.

\section{Comparison of waveforms}
\label{Appendix:gw-wiener-product}

For GW analysis and parameter estimation studies one often assumes Gaussian noise distribution,
which motivates the definition of a detector-noise-weighted inner product
(Wiener product)~\cite{damour2011accuracy,cutler1994gravitational}
to study the space of real time signals $a(t)$ and $b(t)$. It is defined as
\begin{align}
  (a|b)_{S_n, [0,\infty)} = \int_{0}^{\infty} \dd f~\frac{\tilde{a}(f) {\tilde{b}}^\ast(f)}{S_n(f)} \, ,
  \label{eq:wiener-product}
\end{align}
where $\tilde{a}(f)$ and $\tilde{b}(f)$ are the respective
Fourier transforms and $S_{n}(f)$ is the PSD of the noise $n(t)$.
This product defines a norm which is positive-definite and assigns an Euclidean structure
to the vector space of real signals~\cite{damour2011accuracy}.
If one is given two normalized signals $\hat{a}$ and $\hat{b}$,
such that $(\hat{a},\hat{a})_{S_n,[0,\infty)} = 1$ and analogously for $\hat{b}$,
then a result of $(\hat{a},\hat{b})_{S_n,[0,\infty)}$ that is close to 1
indicates that $a$ and $b$ are close to each other.

In practice, one often utilizes the GW strain $h$ together with~\eqref{eq:wiener-product}
for parameter studies.
However, the reconstruction of $h$ from $\Psi_4$ can suffer from
uncertainties and even render an analysis based on $h$ useless.
Recently, in the context of Proca star head-on collision it was shown
in~\cite{bustillo2022gravitational} that an analysis using $\Psi_4$ data
together with~\eqref{eq:wiener-product} and second differenced Gaussian noise PSD
$S_{\Psi_4}$ is equivalent to the more commonly practiced analysis based on $h$ and $S_n$,
while at the same time it removes the uncertainties due to free parameters in the $h$ reconstruction procedure.
For the results presented in the main text we adopted this analysis
which requires additional (but parameter free) processing.
In particular, given two $\Psi_{4,20}(u)$ data streams
we compute~\cite{bustillo2022gravitational}
(in practice we use the extrapolated $r_\infty\PsiFT$ data)
\begin{align}
  \tilde{\Psi}'_{4,20}[k] &= \frac{1 - \cos(2\pi k\Delta f\Delta t)}{2\pi^2(k\Delta f\Delta t)^2}
  \tilde{\Psi}_{4,20}(k \Delta f) \, ,
  \label{eq:2nd-diff-ft-psi4}
  \\
  k &= 0, \dots, N_d-1 \, , \nonumber
\end{align}
where $\tilde{\Psi}_{4,20}(f)$ is the DFT of $\Psi_{4,20}$,
$N_d$ is the length of the output data stream, $\Delta t$ is the time resolution of
$\PsiFT$ and $\Delta f = 1/(N_d\Delta t)$.
This transformation relates the second differenced Fourier transform $\tilde{\Psi}'_{4,20}$
with the Fourier transform of the second derivative $\tilde{\Psi}_{4,20} = \tilde{\ddot{h}}$.
In this work we exclusively used the \texttt{T1800044-v5} Advanced LIGO sensitivity design curve~\cite{ligopsd}
for $S_n$ which is also processed through~\cite{bustillo2022gravitational}
\begin{align}
  S_{\Psi_4}[k] = \frac{1}{(\Delta t)^4} & \left(
    6 - 8 \cos(\frac{2\pi k}{N_d}) + \right. \nonumber \\
    & \left. + 2 \cos(\frac{4\pi k}{N_d})
  \right) S_n[k] \, .
  \label{eq:2nd-diff-psd}
\end{align}
Given these quantities we then compute the Wiener product between
two signals $\Psi^{(1)}_{4,20}$ and $\Psi^{(2)}_{4,20}$ by
\begin{align}
  W(\Psi^{(1)}_{4,20}, \Psi^{(2)}_{4,20}) := (\hat{\Psi}^{(1)}_{4,20}|\hat{\Psi}^{(2)}_{4,20})_{S_{\Psi_4},[f_{\textrm{min}},f_{\textrm{max}}]} \, ,
  \label{eq:wiener-product-psi4}
\end{align}
where we limited the integration to the range
$[f_{\textrm{min}}, f_{\textrm{max}}] = [5~\text{Hz},1000~\text{Hz}]$ for practical purposes,
since the detector PSD is only sampled for a finite frequency range.
The numerical evaluation of the integral is carried out by using a cubic interpolation
to map $\tilde{\Psi}_{4,1}$, $\tilde{\Psi}_{4,2}$ and $S_{\Psi_4}$ to a common frequency grid
and then use a trapezoidal quadrature rule to compute the integral in~\eqref{eq:wiener-product}.
We verified that the results presented in \autoref{fig:wiener-product-vs-n}
are insensitive to the choice of frequency range.
We want to point out that care must be taken when evaluating~
\eqref{eq:2nd-diff-ft-psi4} and~\eqref{eq:2nd-diff-psd}, since a limited frequency range
$[f_{\textrm{min}}, f_{\textrm{max}}]$ also limits the range of $k$ indices to use for
numerical integration, because $f_k = k \Delta f$.

The results in the main text are presented in Planck units and with a particular choice of rescaling
of all quantities (see \aref{Appendix:units}).
Because of that, the obtained results depend on the experimentally unknown scalar-field mass $\mu$.
The choice of $\mu$ influences \eg{} the time scale with which physical processes take place
and, thus, controls the frequency of the GW signals when converted to SI units.
Since we are only interested in directly comparing GWs obtained from
different initial data construction techniques,
we decided to fix $\mu = 1~\times~10^{-11}~\text{eV}$ for all computations.
With this choice the GW's spectrum falls \textit{roughly} into the sensitive region
of the detector for the range of boost parameters $v$ we study,
\ie{} the frequencies of the dominant amplitudes of the PSD of $\PsiFT$ are then $\mathcal{O}(100~\text{Hz})$~\cite{aligo2020}.

\section{Unit systems}
\label{Appendix:units}

Let $l$, $t$, $m$ and $A$ denote length, time, mass and scalar-field amplitude
and let $\hat{l}$, $\hat{t}$, $\hat{m}$ and $\hat{A}$ be the numerical values
given with respect to reference values $l_0$, $t_0$, $m_0$ and $A_0$, \ie{}
\begin{align}
  l &= \hat{l}~l_0 \, ,
  &
  t &= \hat{t}~t_0 \, ,
  &
  m &= \hat{m}~m_0 \, ,
  &
  A &= \hat{A}~A_0 \, .
  \label{eq:rescaled-vars}
\end{align}

In the main text we use the Planck unit system in which $G = c = \hbar = 1$.
From the definition of the Planck length, time, and mass
\begin{align}
  \LPl &= \sqrt{\frac{\hbar G}{c^3}} \, ,
  &
  \TPl &= \sqrt{\frac{\hbar G}{c^5}} \, ,
  &
  \MPl &= \sqrt{\frac{\hbar c}{G}} \, ,
  \label{eq:planck-units}
\end{align}
respectively, we then infer that $\LPl = \TPl = \MPl = 1$ holds in such a system.
Furthermore, all quantities effectively \textit{lose} their physical dimensions
and $l_0$, $t_0$, $m_0$ and $A_0$ now merely serve as dimensionless rescalings.
One possible choice for reference values in which this is realized is
\begin{align}
  l_0 &= 1/\mu \, ,
  &
  t_0 &= 1/\mu \, ,
  &
  A_0 &= 1 \, ,
  \label{eq:base-units-1}
\end{align}
where $\mu$ is the scalar-field mass introduced in the main text,
and we fix $m_0$ further below.
The numerical values of Planck length and time then read $\hLPl = \hTPl = \mu$.
This is a convenient choice, because it eliminates $\mu$ from all equations
without loss of generality.
To see this consider the action~\eqref{eq:action}, which is the fundamental object
of the theory under study.
Using~\eqref{eq:base-units-1} we can rewrite it as
\begin{align}
  S &= \hat{S} S_0 = \frac{1}{\mu^2} \int \dd^4 \hat{x} \sqrt{-g}
      \label{eq:rescale-action}
      \\
      &\times \left(
      \frac{\tensor[^{(4)}]{\hat{R}}{}}{16\pi}
      - \frac{1}{2} \left( g^{ab} \hat{\nabla}_a \hat{\phi}^\ast \hat{\nabla}_b \hat{\phi} + |\hat{\phi}|^2 \right)
  \right) \, , \nonumber
\end{align}
where we omit carets from $\sqrt{-g}$ and $g^{ab}$, since they
are rescaled together with $\dd^4 x$ and $(\nabla_{a} \cdot) (\nabla_{b} \cdot)$, respectively.
Choosing $S_0 = 1/\mu^2$ as the reference value for the action then eliminates all factors of $\mu$,
such that~\eqref{eq:rescale-action} is \textit{formally} equivalent
to~\eqref{eq:action} when also setting $\mu = 1$ in~\eqref{eq:potential-free-scalar},
which was done in the main text.

We note that $m_0$ need not be specified in order to obtain~\eqref{eq:rescale-action}.
Instead, one way to fix it is by demanding
\begin{align}
  1 = \hat{\hbar} S_0 = \hbar =
  \frac{\MPl \LPl^2}{\TPl} = \frac{\hMPl\hLPl^2}{\hTPl} \frac{m_0 l_0^2}{t_0} \, ,
  \label{eq:fix-m0-1}
\end{align}
so that
\begin{align}
  \hat{\hbar} &= \hMPl \mu \, ,
  &
  S_0 &= \frac{1}{\mu^2} = \frac{m_0 l_0^2}{t_0} \, ,
  \label{eq:fix-m0-2}
\end{align}
which implies
\begin{align}
  m_0 &= 1/\mu \, .
  \label{eq:base-units-2}
\end{align}
In summary, \eqref{eq:base-units-1} and~\eqref{eq:base-units-2} are the reference values
one can adopt to eliminate all occurrences of $\mu$ from all equations
when working with Planck units and BSs.

Another set of units commonly found in the literature are the natural units
in which $\hbar = c = 1$.
From~\eqref{eq:planck-units}
one then obtains the relations $\LPl = \TPl = 1/\MPl$ and
the Einstein-Hilbert term in the action also attains an extra factor of $1/G = \MPl^2$.
Thus, when working in such units we recommend the choice
\begin{align}
  l_0 &= 1/\mu \, ,
  &
  t_0 &= 1/\mu \, ,
  &
  A_0 &= \MPl \, ,
  \label{eq:base-units-natural-1}
\end{align}
as reference values.
This then allows to adopt $S_0 = \MPl^2/\mu^2$ as the reference value
for the action so that all occurrences of $\mu$ and $\MPl$ are eliminated from it.
Again, to arrive at $S_0$ we need not specify $m_0$.
Instead the latter is fixed by a similar argument as used for~\eqref{eq:fix-m0-1} and~\eqref{eq:fix-m0-2},
which yields
\begin{align}
  m_0 &= \frac{\MPl^2}{\mu} \, .
  \label{eq:base-units-natural-2}
\end{align}
In summary, \eqref{eq:base-units-natural-1} and~\eqref{eq:base-units-natural-2}
are a convenient choice of reference values one can adopt when working with natural units and BSs
to eliminate all factors of $\mu$ and $\MPl$ from the equations.

We note that~\cite{helfer2022malaise,evstafyeva2023boson} also works with natural units
but uses
\begin{align}
  l_0 &= 1/\mu \, ,
  &
  t_0 &= 1/\mu \, ,
  &
  m_0 &= 1/\mu \, ,
  &
  A_0 &= \MPl \, .
  \label{eq:base-units-natural-ref}
\end{align}
This set of reference values also removes all factors of $\mu$ and $\MPl$ from the action.
However, we find that this choice is not compatible with other parts of our analysis.
As an example consider the $g_{rr}$ component in~
\eqref{eq:ansatz-metric-1d} which,
when expressed in natural units and non-rescaled variables,
must read
\begin{align}
  \frac{1}{g_{rr}} = 1 - \frac{2Gm}{r} = 1 - \frac{2m}{\MPl^2 r} \, ,
\end{align}
since $g_{rr}$ is dimensionless.
Performing a rescaling using our convention~\eqref{eq:base-units-natural-1}
and~\eqref{eq:base-units-natural-2} eliminates $\MPl$,
whereas using~\eqref{eq:base-units-natural-ref} does not.

Yet another set of units that is commonly used are geometric units in which $G = c = 1$.
From~\eqref{eq:planck-units} one obtains $\LPl = \TPl = \MPl$ and
now only the term involving the scalar-field potential in the action attains
an extra factor of $1/\hbar^2 = 1/\MPl^4$. For this setup we recommend the choice
\begin{align}
  l_0 &= \MPl^2/\mu \, ,
  &
  l_0 &= \MPl^2/\mu \, ,
  &
  A_0 &= 1 \, .
  \label{eq:base-units-geometric-1}
\end{align}
This then allows to adopt $S_0 = \MPl^4 / \mu^2$ to eliminate all factors of $\mu$ and $\MPl$
from the action.
By a similar argument we used before, we fix now $m_0$ through
\begin{align}
  \MPl^2 = \hbar = \hat{\hbar} S_0 = \hat{\hbar} \frac{m_0 l_0^2}{t_0} \, ,
\end{align}
and find
\begin{align}
  m_0 = \MPl^2 / \mu \, .
  \label{eq:base-units-geometric-2}
\end{align}
In summary, \eqref{eq:base-units-geometric-1} and~\eqref{eq:base-units-geometric-2}
are a convenient choice of reference values one can adopt when working with geometric units and BSs
to eliminate all factors of $\mu$ and $\MPl$ from the equations.

\section{Construction of isolated boson stars in 1D}
\label{Appendix:isolated-bs}

In this appendix we fix a typo from~\cite{helfer2022malaise}
in the equations that describe stationary, spherically symmetric and isolated BS models.
The starting point for the construction of such stars are
ans\"atze for the metric and scalar field which are of the form
\begin{align}
  \dd s^2 &= - e^{2\Phi} \dd t^2 + \left( 1 - \frac{2m}{r} \right)^{-1} \dd r^2
  \nonumber\\
  &+ r^2 \left( \dd \theta^2 + \sin(\theta)^2 \dd \varphi^2 \right) \, ,
  \label{eq:ansatz-metric-1d}
  \\
  \phi(t,r) &= A(r) e^{i\omega t} \, ,
\end{align}
where $A, \omega$ are the amplitude and harmonic angular frequency of the scalar field.
Inserting these into the EKG system yields
\begin{align}
  \partial_r \Phi &= \frac{m}{r(r-2m)}
  \nonumber\\
  &+ \frac{2\pi r^2}{r-2m}
  \left( \eta^2 + \omega^2 e^{-2\Phi} A^2 - V \right) \, ,
  \label{eq:typo1}
  \\
  \partial_r m &= 2\pi r^2 \left( \eta^2 + \omega^2 e^{-2\Phi} A^2 + V \right) \, ,
  \\
  \partial_r A &= \left( 1 - \frac{2m}{r} \right)^{-1/2} \eta \, ,
  \\
  \partial_r \eta &= - 2 \frac{\eta}{r} - \eta \partial_r \Phi
  \nonumber\\
  &+ \left( 1 - \frac{2m}{r} \right)^{-1/2}
  \left( V' - \omega^2 e^{-2\Phi} \right) A \, ,
\end{align}
where $V' = \dd V / \dd A^2$.
In~\cite{helfer2022malaise} a factor $r/(r - 2m)$
was missing in the second term on the RHS in~\eqref{eq:typo1}.

This system of equation is subjected to the boundary conditions
\begin{align}
  A(0) &= A_{\textrm{ctr}} \, ,
  &
  m(0) &= 0 \, ,
  &
  \eta(0) &= 0 \, ,
  \\
  \partial_r \Phi(0) &= 0 \, ,
  &
  \lim_{r\to\infty} A(r) &\to 0 \, .
\end{align}
The asymptotic behavior of $A$ is given by
\begin{align}
  A \sim \frac{1}{r^{1+\epsilon}} e^{-r \sqrt{1 - \omega^2 e^{-2\Phi}}} \, ,
  \label{eq:asymp-phi}
\end{align}
where $\epsilon$ is a non-integer correction that was already reported in~\cite{kaup1968klein}
and recently discussed in~\cite{evstafyeva2023boson}.

Our numerical implementation uses the same shooting algorithm as~\cite{helfer2022malaise},
where integration starts at $r = 0$ and proceeds radially outwards.
The asymptotic behavior is used for patching the solution at a
radius $r_m$, determined dynamically during integration, beyond which
the scalar field is no longer integrated, but instead frozen to the behavior~\eqref{eq:asymp-phi}.
Results presented in this work do not account for the $\epsilon$ correction,
as we were simply not aware of it when doing the analysis.
However, we implemented the correction in hindsight and found that
our single star solutions are only altered in $A$ and $\eta$ in the far-field regime
by $\mathcal{O}(10^{-10})$, which is negligible when compared to other error sources.

\section{Conformal metric and its time derivative}
\label{Appendix:notation-conformal-metric}

We note a difference in notation between~\cite{east2012conformal} and our work
which is mainly based on~\cite{baumgarte2010numerical}.
In particular, \cite{baumgarte2010numerical}~defines the time derivative of
the conformal metric $\bar\gamma_{ij}$ as
\begin{align}
  \bar{u}_{ij} := \partial_t \bar{\gamma}_{ij} = - \bar{\gamma}_{ik} \bar{\gamma}_{jl} \partial_t \bar{\gamma}^{kl} \, .
  \label{eq:definition-ubar}
\end{align}
Its indices are also raised with the conformal metric,
\begin{align}
  \bar{u}^{mn} = \bar{\gamma}^{im} \bar{\gamma}^{jn} \bar{u}_{ij} \, .
\end{align}
Using~\eqref{eq:definition-ubar} we obtain
\begin{align}
  \bar{u}^{mn} = - \partial_t \bar{\gamma}^{mn} \, .
  \label{eq:definition-ubaruu}
\end{align}

On the other hand, \cite{east2012conformal} does not work with~\eqref{eq:definition-ubaruu}, but
instead uses
\begin{align}
  \dot{\bar{\gamma}}_{ij} &:= \partial_t \bar{\gamma}_{ij} \, ,
  &
  \dot{\bar{\gamma}}^{ij} &:= \partial_t \bar{\gamma}^{ij} \, .
\end{align}
To make the connection with their notation we had to set
$\bar{u}^{mn} = - \dot{\bar{\gamma}}^{mn}$, which explains the change in sign
in~\eqref{eq:ubar}.

\section{Moving average}
\label{subsection:moving-average}

The differences between $\PsiFT$ obtained from different resolutions $n$ are postprocessed
for the purpose of demonstrating self-convergence in \autoref{fig:psi4_self-convergence}.
To this end, the data is interpolated to a common time grid using cubic splines and
$N = 2000$ points, after which differences between consecutive resolutions are computed.
To suppress the artificial noise due to zero crossings introduced by interpolation errors,
and which are amplified on a logarithmic scale, we apply a moving average filter to
these differences. In particular, let $\Delta \Psi_{4,i}, i = 1,\dots,N$ be the difference
between two interpolated data streams and let $w$ be an averaging window width, assumed
to be even and positive. We define the moving average as
\begin{align*}
  \text{avg}[\Delta \Psi_{4,i}]
  &= \frac{1}{w}\sum_{j=1}^{w} \Delta \Psi_{4,i-w/2+j} \, ,
  &
  i &\in \left[1+\frac{w}{2}, N - \frac{w}{2}\right] \, ,
  \\
  \text{avg}[\Delta \Psi_{4,i}] &= \text{avg}[\Delta \Psi_{4,w/2}] \, ,
  &
  i &\in \left[1, \dots, \frac{w}{2}\right] \, ,
  \\
  \text{avg}[\Delta \Psi_{4,i}] &= \text{avg}[\Delta \Psi_{4,N-w/2}] \, ,
  &
  i &\in \left[N-\frac{w}{2}+1, N\right] \, .
\end{align*}
For the data presented in this work we chose $w = 20$.

\bibliography{paper.bbl}
\end{document}